\documentclass[reprint,superscriptaddress,showpacs,amsmath,amssymb,aip]{revtex4-1}
\usepackage{dcolumn}
\usepackage{amsmath}
\usepackage{amssymb}
\usepackage{amsfonts}
\usepackage{graphicx}
\usepackage{subfigure}
\usepackage{dcolumn}
\usepackage{ulem}
\usepackage[utf8]{inputenc}
\usepackage[T1]{fontenc}
\usepackage{mathptmx}
\usepackage{mathrsfs}
\usepackage{booktabs}
\usepackage{array}
\usepackage{makecell}
\usepackage[pdfpagemode=UseNone,pdfstartview=FitH,colorlinks=true,linkcolor=blue,urlcolor=blue,anchorcolor=blue,citecolor=blue]{hyperref}
\allowdisplaybreaks[4]

\begin{document}
	
	\title{Dissipative couplings in cavity magnonics }
	
	\author{Yi-Pu Wang}
	\email{Yipu.Wang@umanitoba.ca}
	\affiliation{Department of Physics and Astronomy, University of Manitoba, Winnipeg, Canada R3T 2N2}

	\author{Can-Ming Hu}
	\email{hu@physics.umanitoba.ca}
	\affiliation{Department of Physics and Astronomy, University of Manitoba, Winnipeg, Canada R3T 2N2}

	\date{\today}
	\begin{abstract}
	Cavity Magnonics is an emerging field that studies the strong coupling between cavity photons and collective spin excitations such as magnons. This rapidly developing field connects some of the most exciting branches of modern physics, such as quantum information and quantum optics, with one of the oldest science on the earth, the magnetism. The past few years have seen a steady stream of exciting experiments that demonstrate novel magnon-based transducers and memories. Most of such cavity magnonic devices rely on coherent coupling that stems from direct dipole-dipole interaction. Recently, a distinct dissipative magnon-photon coupling was discovered. In contrast to coherent coupling that leads to level repulsion between hybridized modes, dissipative coupling results in level attraction. It opens an avenue for engineering and harnessing losses in hybrid systems.  This article gives a brief review of this new frontier. Experimental observations of level attraction are reviewed. Different microscopic mechanisms are compared. Based on such experimental and theoretical reviews, we present an outlook for developing open cavity systems by engineering and harnessing dissipative couplings.
	\end{abstract}

	\maketitle

\section{Introduction}\label{sec1}
	Coherent coupling is ubiquitous in nature, and it underpins a myriad of applications~\cite{Zoller-09,Kurizki-15,Xiang-13}. Intuitively, coherently coupled subsystems exchange energy between each other periodically with a rate of $J$, which is determined by the coupling strength. The most concerned and explored issue is coherent light-matter interaction~\cite{Cohen-92}. On the one hand, the electromagnetic field is an excellent information carrier because of its fast propagation speed and low loss during long-distance transfer. On the other hand, information processing and storage are mainly based on solid-state devices, such as superconducting qubits~\cite{Devoret-13,You-11}, magnetic materials, spin ensembles~\cite{Fert-88,Kubo-18,Zhu-11,Wu-10}, trapped ions~\cite{Cirac-95,Blatt-05,Bohnet-16}, and semiconductor quantum dots~\cite{Stockklauser-17,Watson-18}, etc.. Hence, coherent coupling based on light-matter interactions plays a crucial role in information processing technology.  It enables coherent information exchange between the information carrier and the processor.
	
\begin{figure}[!t]
		\centering
		\includegraphics[width=0.46\textwidth]{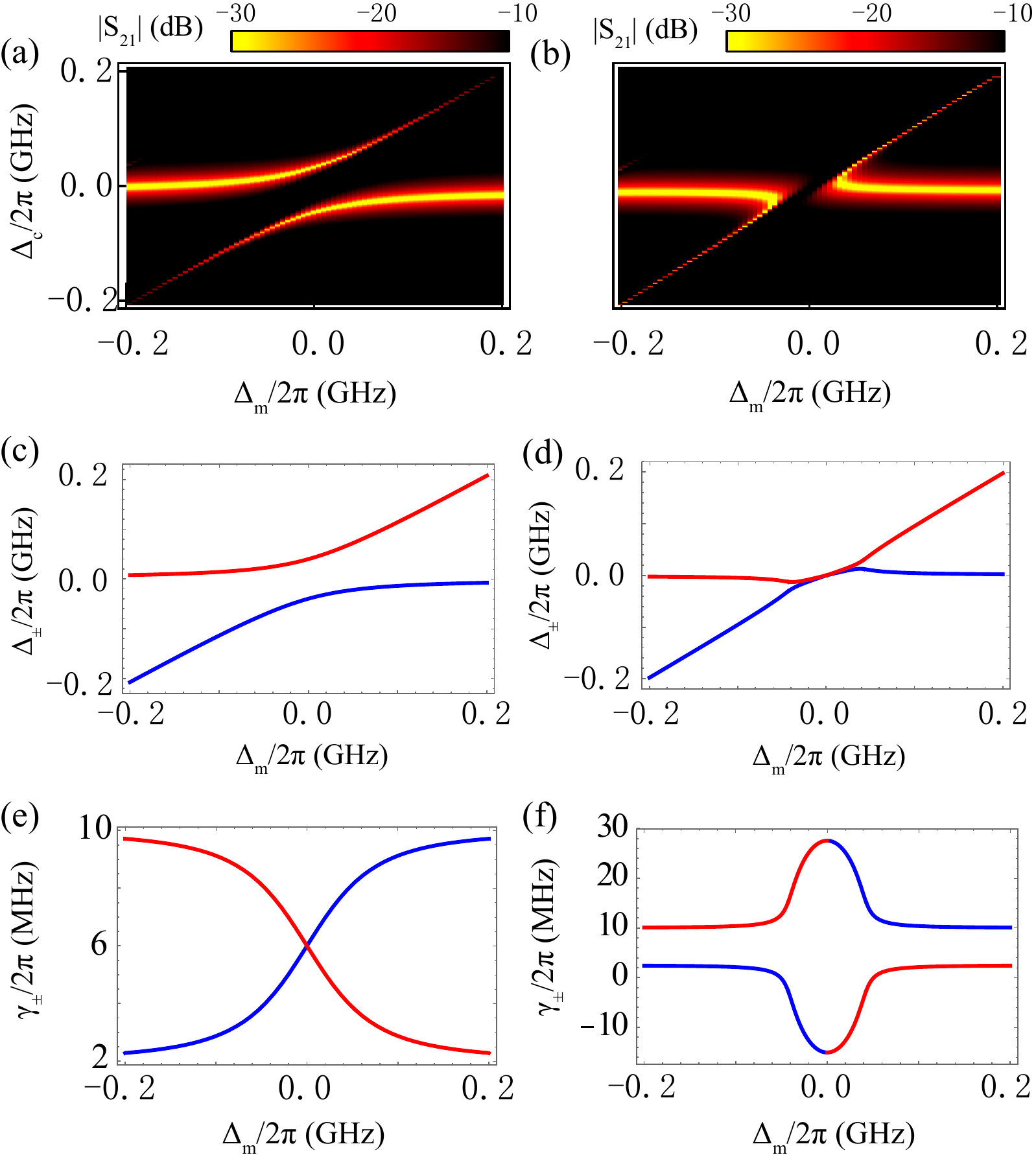}
		\caption{(color online). (a)~Transmission spectrum of a coherently coupled cavity photon-magnon system as a function of the field detuning $\Delta_{m}$ ($\Delta_{m}=\omega_{m}-\omega_{c}$) and the probe field frequency detuning $\Delta_{c}$ ($\Delta_{c}=\omega-\omega_{c}$), which is shown as level repulsion. (b)~Transimission spectrum of a dissipatively coupled cavity photon-magnon system, which is shown as level attraction. (c)~Dispersion of the coherently coupled system. (d)~Dispersion of the dissipatively coupled system. (e)~Linewidths of the hybridized modes of the coherently coupled system as a function of the field detuning $\Delta_{m}$. (f)~Linewidths of the hybridized modes of the dissipatively coupled system as a function of the field detuning $\Delta_{m}$. (Magnon mode frequency :$\omega_{m}$, Cavity mode frequency : $\omega_{c}$). The uncoupled modes' intrinsic damping rates employed here are 10 MHz and 2 MHz for the cavity mode and magnon mode, respectively. The coherent and dissipative coupling strengths both are 40 MHz. (a) and (b) are adapted from Harder \textit{ et al.} \cite{Harder-18}.  }
		\label{fig:1}
	\end{figure}

	
	Recently, a hybrid system based on coherently coupled cavity photons and collective spin excitations (magnons) has attracted much attention~\cite{Huebl-13,Tabuchi-14,Zhang-14,Tobar-14,Hu-15,Maier-16,Cao-15,Ferguson-15,TangJAP-16,Tobarapl-16,Bauer-15,Hu-17,Dengke-17,Lambert-16,TabuchiScience-15,NakamuraSA-17,Osada-16,Zhangxu-16,haighprl-16,Hisatomi-16,Braggio-17,TangNC-15,TangSA-16,Lijie-18,Dany-19,Harder-18s,Hu-16C,Hu-19V,Wang-18,RaoNC-19}. The coherent coupling here relies on the magnetic dipole-dipole interaction between the magnetostatic modes~\cite{Stancil} and cavity photon modes, which increases with increasing spin numbers and cavity mode density. Most experiments have been performed by using the single crystal yttrium iron garnet (YIG) that has a large spin density of about $10^{22}~\mu_{B}~\rm{cm}^{-3}$ ($\mu_{B}$ is the Bohr magneton). Strong coherent coupling is readily realized between the magnetostatic magnons and cavity photons when the ferrite is placed at the magnetic field antinodes of microwave cavities.  The hallmark of coherent magnon-photon coupling is the normal-mode splitting, where the energy levels of the hybridized modes repulse with each other, as shown in Figs.~\ref{fig:1}(a) and 1(c) for the transmission spectrum and dispersion curve, respectively. The past few years have seen a steady stream of exciting experiments that studied and utilized such couplings, which has led to the emergence of cavity magnonics. It is anticipated that this new frontier may have broad impact on quantum information~\cite{Kimble-08,NakamuraSA-17}, spintronics~\cite{Hu-15,Hu-17}, cavity optomagnonics\cite{Osada-16,Zhangxu-16,haighprl-16,Hisatomi-16,Braggio-17}, and quantum magnonics~\cite{TabuchiScience-15,NakamuraSA-17}. In particular, a versatile magnon-based quantum information processing platform has taken shape with coherently coupled microwave photons, magnons, optical photons, phonons, and superconducting qubits.
	
	So far, most of the cavity magnonic systems are based on microwave cavities that have nearly closed boundaries. Measurements on such nearly closed systems are typically done by coupling them weakly to external ports so that the linear response of the systems can be perturbatively measured via microwave transmission or reflection spectroscopy. In such setups, the external dissipation rates (ports to the environment) and the intrinsic damping rates of the coupled modes are relatively small, and the dissipation processes are not obviously correlated. Then coherent coupling is the dominate coupling mechanism in such systems. However, in 2018, a distinct magnon-photon dissipative coupling was discovered \cite{Harder-18}, and it has been quickly verified in a variety of setups with different cavity configurations~\cite{Bhoi-19,Ying-19,Rao-19,PC-19,XiaB-19,Wang-19,Rao-19-2,Yao-19,Yu-19,Du-19,YuYang-19,Yao-19-2}. A distinct feature of a dissipatively coupled system is the level attraction (LA) of the hybridized modes as shown in Figs.~\ref{fig:1}(b) and 1(d), which is in strong contrast to the level repulsion induced by coherent coupling. It should be noted that apart from the cause of the dissipative coupling, the LA effect itself is a ubiquitous phenomenon that can also be induced by many other mechanisms, such as two-tone driving \cite{Xia-18,Boventer-19,Boventer-19-2}, parametric driving \cite{Bernier-17,Igor-19}, and coupled systems involving inverted oscillators~\cite{Kohler-18}. However, as we will explain in this article, the dissipative coupling that can be detected via the effect of LA is of particular interest for developing cavity magnonics.
	
	Initially, the physics of dissipative magnon-photon coupling is understood by the classical picture of the cavity Lenz effect (see section \ref{sec3A}) \cite{Harder-18}. Via a series of recent studies\cite{Wang-19,Rao-19-2,Yao-19,Yu-19,Du-19,Yao-19-2}, its quantum mechanical origin is now fully revealed, which shows that it is an indirect coupling mechanism mediated by dissipative channels in open cavity systems. Specifically, the external dissipations of magnons and photons are non-trivial but correlated due to their interactions with the common reservoir. Interestingly enough, with the dissipative coupling in such open systems, the loss is no longer a nuisance, but becomes a valuable means for controlling the system \cite{Clerk-15} to enable new applications. In particular, it allows the control of the arrow of time via engineered dissipation, which is demonstrated in a recent experiment that measures nonreciprocal microwave transmissions in cavity magnonics\cite{Wang-19}. Furthermore, it enables connecting distant oscillators via nonlocal dissipative interactions~\cite{PC-19,Du-19,XiaB-19,Rao-19-2}.  This opens a new path for developing cavity magnonics to make devices with unique functionalities such as backscattering isolation, nonreciprocal wave propagation, and unidirectional signal amplification~\cite{Clerk-15,Caloz-18,Chapman-17,Bernier-17,Peterson-17,Lecocq-17,Fink-17}. All of these may lead to hybrid systems with improved operational efficiency, or sensors with enhanced sensitivity.
		
This article is split into four main sections. Following the introduction, we first provide in section \ref{Sec2} a brief review of recent experiments on dissipative magnon-photon coupling and level attraction, where we distinguish them according to different mechanisms. Then in section \ref{Theory}, we discuss several different theoretical pictures, with the focus on comparing and clarifying the underlying mechanism of the dissipative coupling and the level attraction. Finally, we present in section \ref{sec4} an outlook for developing open cavity systems that may harness losses via the engineered dissipative coupling.

\section{Experiments on dissipative magnon-photon coupling and level attraction}\label{Sec2}
Recent experiments that observed magnon-photon level attraction are summarized in Table~\ref{Table1}. Our discussion here is focused on level attraction induced by dissipation magnon-photon coupling (Sec. \ref{sec2A}). Other mechanisms that can also induce level attraction, such as two-tone driven (Sec. \ref{sec2B}) and indirect coupling between magnon modes in multi-YIG spheres (Sec. \ref{sec2C}), are only briefly introduced.
\begin{table*}[!t]
	\def\arraystretch{1}
	\caption{\footnotesize{A summary of key experimental devices and setups of dissipative magnon-photon coupling and level attraction. In Refs.~\cite{Harder-18,Yao-19,Yao-19-2,Bhoi-19,Ying-19,Wang-19,Rao-19} the dissipative coupling was due to dissipation processes, while in Ref. \cite{Boventer-19,Boventer-19-2} the level attraction was caused by the interference effect between two driven tones. The figure in the third row is adapted from Wang \textit{ et al.} \cite{Wang-19}. The figure in the fourth row is adapted from Rao \textit{ et al.} \cite{Rao-19}.}}
	\centering
	\begin{tabular}{>{\centering\arraybackslash}m{4cm}>{\centering\arraybackslash}m{6.2cm}>{\centering\arraybackslash}m{4.8cm}}
		\toprule
		\multicolumn{2}{l}{{\bf Dissipation induced dissipative coupling}}
		\\ \toprule
		Reference & Device and setup & Key features \\ \midrule \\
		\begin{center}
			Harder \textit{et al.}\cite{Harder-18}\\Yao and Yu \textit{et al.} \cite{Yao-19,Yao-19-2}
		\end{center} & \includegraphics[width=4.8cm]{Table1Fig1.pdf} & \begin{center}
		\begin{flushleft}
				${\small\bullet}$~1D Fabry-Perot-like cavity\\
		\end{flushleft}\begin{flushleft}
		${\small\bullet}$~Level attraction\\
	\end{flushleft}\begin{flushleft}
	${\small\bullet}$~$\kappa/2\pi$= 112 MHz
\end{flushleft}
		\end{center}\\ \\
		\begin{center}
			Bhoi \textit{et al.} \cite{Bhoi-19}
		\end{center}
		&
		\includegraphics[width=3.9cm]{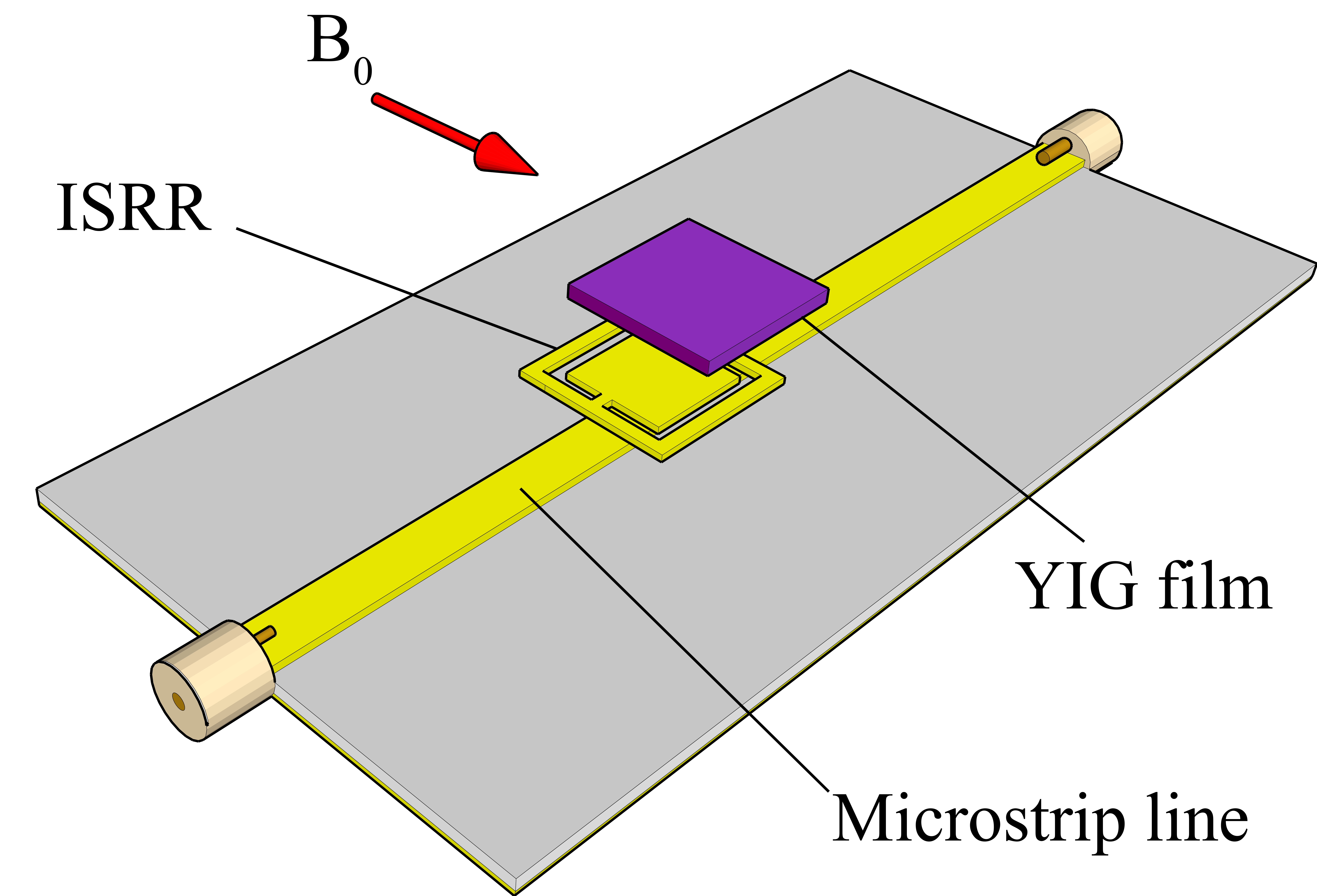} &  \begin{center}
			\begin{flushleft}
				${\small\bullet}$~Inverted pattern of the split-ring resonator\\
			\end{flushleft}\begin{flushleft}
			${\small\bullet}$~Level attraction
		\end{flushleft}
		\end{center}\\ \\
		\begin{center}
			Yang \textit{et al.} \cite{Ying-19}\\Wang \textit{et al.} \cite{Wang-19}
		\end{center} & \includegraphics[width=4cm]{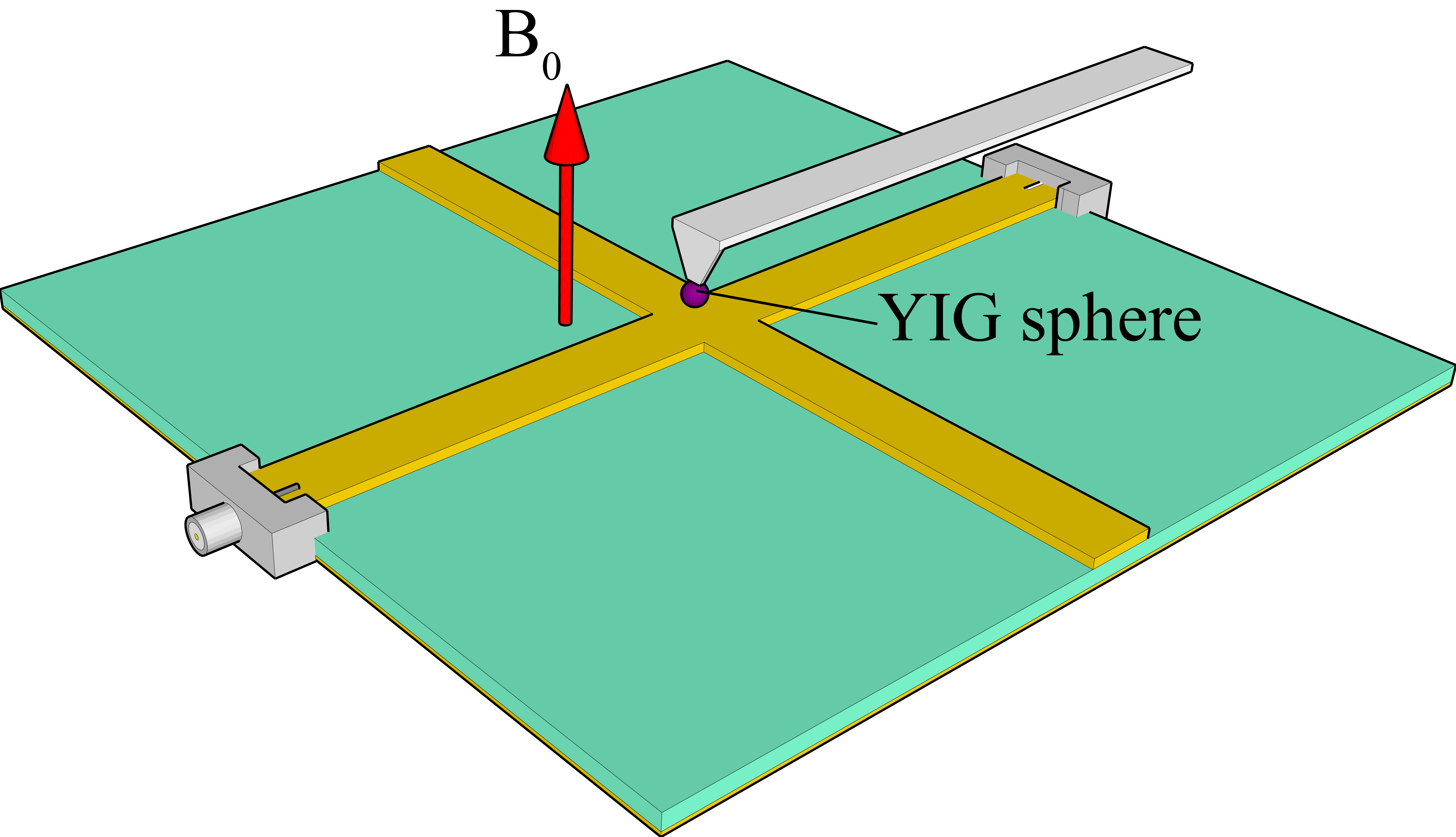} & \begin{center}
			\begin{flushleft}
				${\small\bullet}$~Cross-line microwave circuit\\
			\end{flushleft}\begin{flushleft}
			${\small\bullet}$~Level attraction\\
		\end{flushleft}\begin{flushleft}
		${\small\bullet}$~Non-reciprocial microwave propagation\\
	\end{flushleft}\begin{flushleft}
	${\small\bullet}$~$\kappa/2\pi$= 880 MHz
\end{flushleft}
		\end{center}\\ \\
		\begin{center}
			Rao \textit{et al.} \cite{Rao-19}
		\end{center}&
		\includegraphics[width=3.5cm]{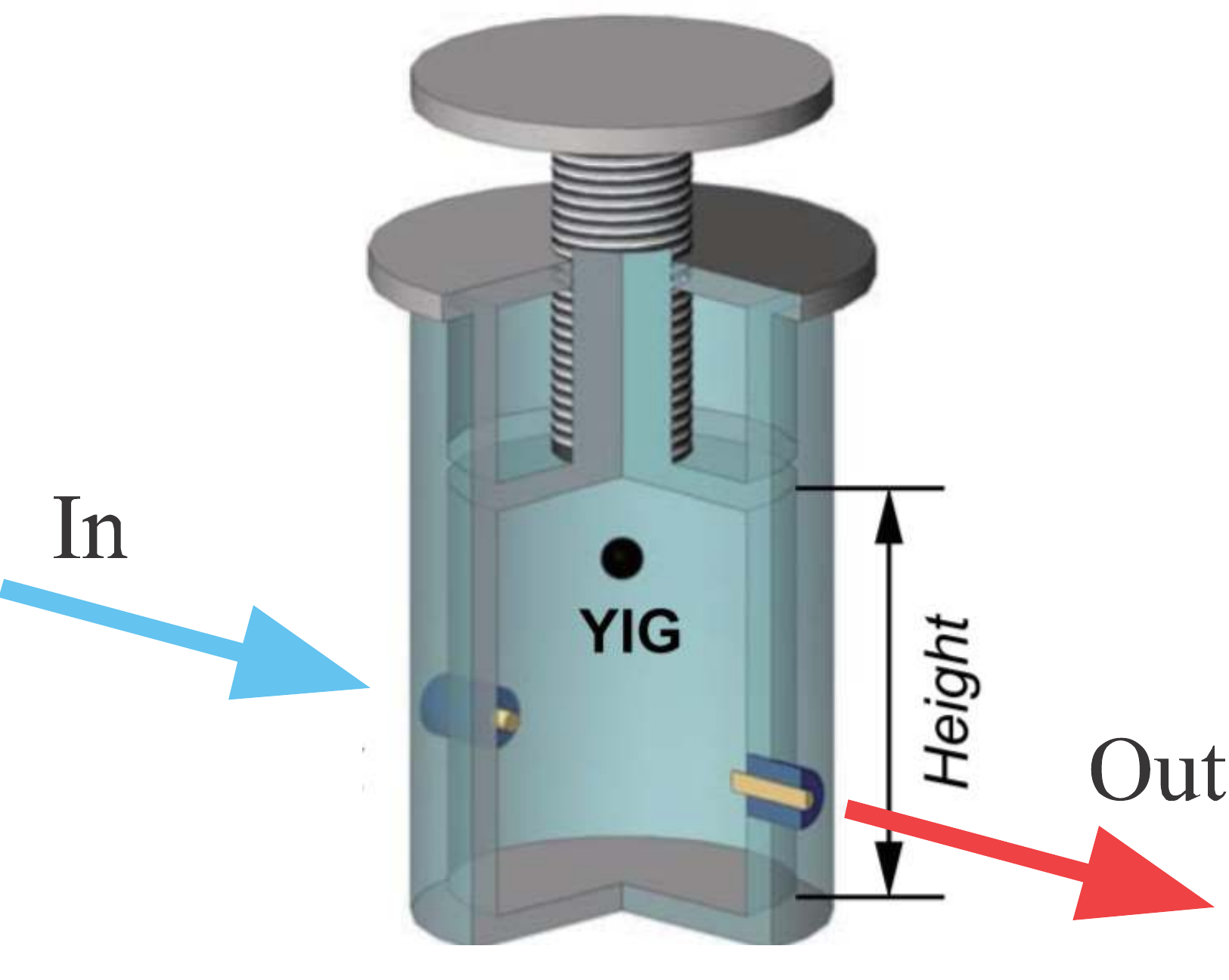} &  \begin{center}
			\begin{flushleft}
				${\small\bullet}$~Anti-resonance within a 3D cavity\\
			\end{flushleft}\begin{flushleft}
			${\small\bullet}$~Level attraction\\
		\end{flushleft}\begin{flushleft}
		${\small\bullet}$~$\kappa/2\pi$= 14.99 GHz
	\end{flushleft}
		\end{center}\\ \\
		\toprule
		\multicolumn{2}{l}{{\bf Two-tone driven experiment
			}} \\ \toprule
		Reference & Experimental setup & Key features \\ \midrule \\
		\begin{center}
			Boventer \textit{et al.} \cite{Boventer-19,Boventer-19-2}
		\end{center} &\includegraphics[width=5.4cm]{Table1Fig5.pdf} &  \begin{center}
			\begin{flushleft}
				${\small\bullet}$~Two driven tones\\
			\end{flushleft}\begin{flushleft}
			${\small\bullet}$~`Cavity' port and `Magnon' port\\
		\end{flushleft}\begin{flushleft}
		${\small\bullet}$~Phase shifter
	\end{flushleft}
		\end{center}\\
		\bottomrule
	\end{tabular}
	\label{Table1}
\end{table*}

\subsection{Level attraction between cavity anti-resonance and magnon mode}\label{sec2A}
Level attraction induced by the dissipative magnon-photon coupling was discovered by Harder $et~al$. in 2018~\cite{Harder-18}. The experiment was performed by placing a YIG sphere in an open cavity, as shown in the first row of Table~\ref{Table1}. The open cavity is made of a 1D Fabry-Perot-like cavity, where a circular waveguide is connected to circular-rectangular transitions. It forms cavity modes but also allows transmission of a traveling wave. Magnon-photon level repulsion and level attraction are observed, when the YIG sphere is placed at the antinode and node of the cavity magnetic field, respectively. The experiment reveals both coherent and dissipative magnon-photon couplings, which can be controlled by changing the YIG location in the cavity.

A similar level attraction (abnormal anticrossing) was experimentally observed by Bhoi $et~al$~\cite{Bhoi-19} using a planar geometry system that consists of a YIG film and an inverted pattern of split-ring resonator (ISRR) structure. In their set up, both of the YIG sample and ISRR are coupled to a microstrip feeding line, as shown in Table~\ref{Table1}, which also feeds in a traveling wave through the coupled system.

Level attraction in a different planar geometry system was demonstrated by Yang $et~al$ ~\cite{Ying-19}, where a planar microwave cross-line circuit was coupled with YIG sphere~\cite{Ying-19}. This microwave circuit consists of an open-ended microstrip resonator galvanically connected to a transmission line at its crossing center, as shown in the third row of Table~\ref{Table1}. The YIG sphere couples with both of the electromagnetic fields of the microstrip resonator and the transmission line. Both the coherent and dissipative couplings between the cavity and magnon modes are achieved in this planar device, which can be controlled by adjusting the YIG position.

The experiments summarized above have one common feature. That is, all of the devices support both cavity mode (standing wave) and traveling wave propagation. The interference between the standing and traveling waves leads to a cavity anti-resonance, which appears as a dip with a broad background in the transmission spectra. This is the general feature of waveguides galvanically connected with resonant structures, where the resonant structure can be an atomic system, a superconducting qubit, or harmonic oscillators.  Interestingly, Rao $et~al$~\cite{Rao-19} have found that even in the conventional 3D microwave cavity, as shown in Table~\ref{Table1}, the anti-resonance can be formed with specific configuration of the cavity input and output ports, which enables additional dissipation of the cavity photons to the external environment.

In addition to the intrinsic damping of the cavity modes formed by the standing waves, anti-resonances are further characterized by an external damping rate $\kappa$, which describes the radiative coupling between the cavity photon and the traveling wave. The input-output relation has the general from~\cite{Aspelmeyer-14}:
\begin{equation}
\hat{p}_{out}=\hat{p}_{in}-\sqrt{\kappa}\hat{a}, \label{eq:input-output}
\end{equation}
where $\hat{p}_{in(out)}$ is the input (output) field and $\hat{a}$ is the cavity mode photon operator. Without the cavity mode, Eq.~(\ref{eq:input-output}) describes the traveling wave that propagates from the input to the outport port ($\hat{p}_{out}=\hat{p}_{in}$).

The external damping rates $\kappa$ deduced from the cavity anti-resonances in different cavity circuits are listed in Table~\ref{Table1}. In Sec.~\ref{Theory}, we will see that the magnon-photon dissipative coupling strength $\Gamma$ is simply determined by the square root of the product of the external damping rates of the magnons and cavity photons. The external damping rate of the magnons in the YIG sphere is quite small due to its weak magnetic dipole interaction with the open environment. Hence, in order to obtain a large dissipative coupling strength, a relatively large external damping rate $\kappa$ is required for the cavity mode. This sets in two conditions for observing level attraction caused by the magnon-photon dissipative coupling: a traveling wave and a large external damping rate of the cavity photons.

In addition to the dissipative coupling, conventional magnon-photon coherent coupling coexists in these open cavity systems, which arises from the overlapping between the ferrite and the microwave magnetic field of the cavity mode. The interplay of coherent and dissipative coupling is of nontrivial consequence, leading to nonreciprocal microwave transmission and unidirectional invisibility, which was discovered very recently by Wang $et~al.$ \cite{Wang-19}.

\subsection{Level attraction with two-tone driven scheme}\label{sec2B}
Observing level attraction does not always indicate dissipative magnon-photon coupling. For example, a two-tone driven scheme can also lead to level attraction in a coupled magnon-photon system, which was proposed by Grigoryan $et~al.$\cite{Xia-18} and first measured by Boventer $et~al.$~\cite{Boventer-19,Boventer-19-2}. The key feature of the experimental setup is that the drive field is split into two paths, one is applied to the cavity input port, and the other one is applied through a loop antenna directly to the YIG sample as depicted in Table~\ref{Table1}. The anticrossing gap at resonance can be set to zero by merely controlling the amplitudes and relative phase of the inputs. The linewidths of the hybridized modes and the effective coupling strength were systematically adjusted and analyzed by Boventer $et~al.$~\cite{Boventer-19,Boventer-19-2}.

\subsection{Dissipative coupling between multi-YIG spheres}\label{sec2C}
Furthermore, an effective dissipative coupling between two magnon modes in two spatially separated YIG spheres has been achieved by coupling them to the same cavity mode~\cite{PC-19,XiaB-19}. As demonstrated by Xu $et~al.$\cite{PC-19}, if one of the magnon modes dissipatively coupled with the cavity mode and the other one coherently coupled with the same cavity mode, then the effective coupling between these two magnon modes is a dissipative one. However, if both of the magnon modes are coherently or dissipatively coupled to the cavity mode, then the effective long-range coupling is a coherent one.

\begin{table*}[!t]
	\def\arraystretch{1}
	\caption{\footnotesize{A summary of several theoretical pictures of the dissipative coupling and level attraction. The figure in the third row is adapted from Yao \textit{ et al.} \cite{Yao-19}.}}
	\centering
	\begin{tabular}{>{\centering\arraybackslash}m{4cm}>{\centering\arraybackslash}m{6cm}>{\centering\arraybackslash}m{4.8cm}}
		\toprule
		\multicolumn{2}{l}{{\bf Comparison of different theoretical pictures}}
		\\ \toprule
		Reference & Schematic diagram & Key features \\ \midrule \\
		\begin{center}
			Harder \textit{et al.}
			\cite{Harder-18}\end{center} & \includegraphics[width=3.8cm]{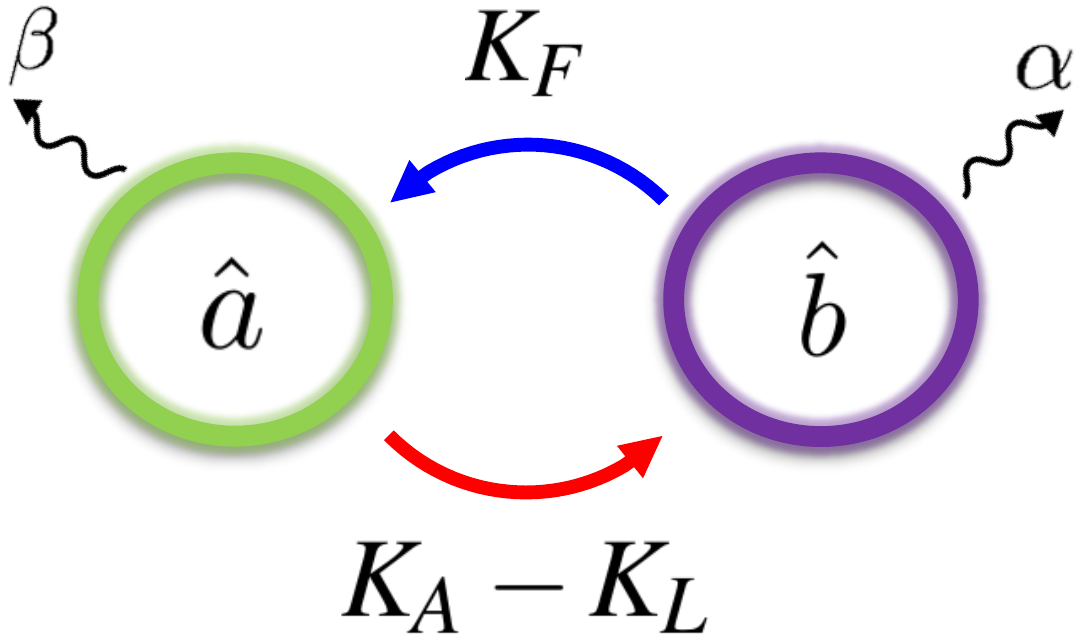} & \begin{center}
			\begin{flushleft}
				${\small\bullet}$~Cavity Lenz effect\\
			\end{flushleft}\begin{flushleft}
			${\small\bullet}$~Dissipative coupling\\
		\end{flushleft}\begin{flushleft}
		${\small\bullet}$~Level attraction
	\end{flushleft}
		\end{center}\\ \\
		\begin{center}
			Wang \textit{et al.} \cite{Wang-19}\\Rao \textit{et al.} \cite{Rao-19-2}\\Metelmann \textit{et al.} \cite{Clerk-15}
		\end{center} & \includegraphics[width=4cm]{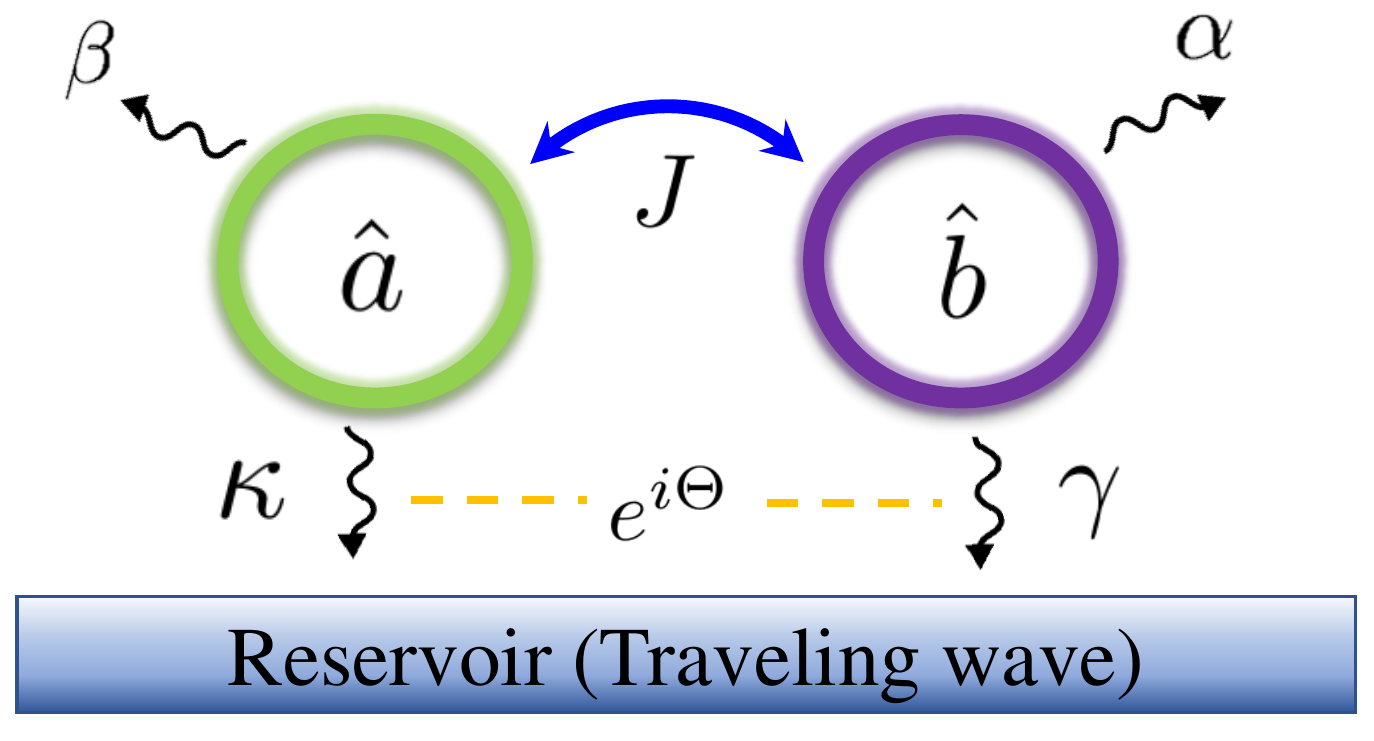} & \begin{center}
			\begin{flushleft}
				${\small\bullet}$~Traveling-wave-type reservoir \\
			\end{flushleft}\begin{flushleft}
			${\small\bullet}$~Cooperative and correlated dissipations\\
		\end{flushleft}\begin{flushleft}
		${\small\bullet}$~Dissipative coupling\\
	\end{flushleft}\begin{flushleft}
	${\small\bullet}$~Level attraction
\end{flushleft}
		\end{center}\\ \\
		\begin{center}
			Yao and Yu \textit{et al.} \cite{Yao-19,Yao-19-2}
		\end{center}&
		\includegraphics[width=3.8cm]{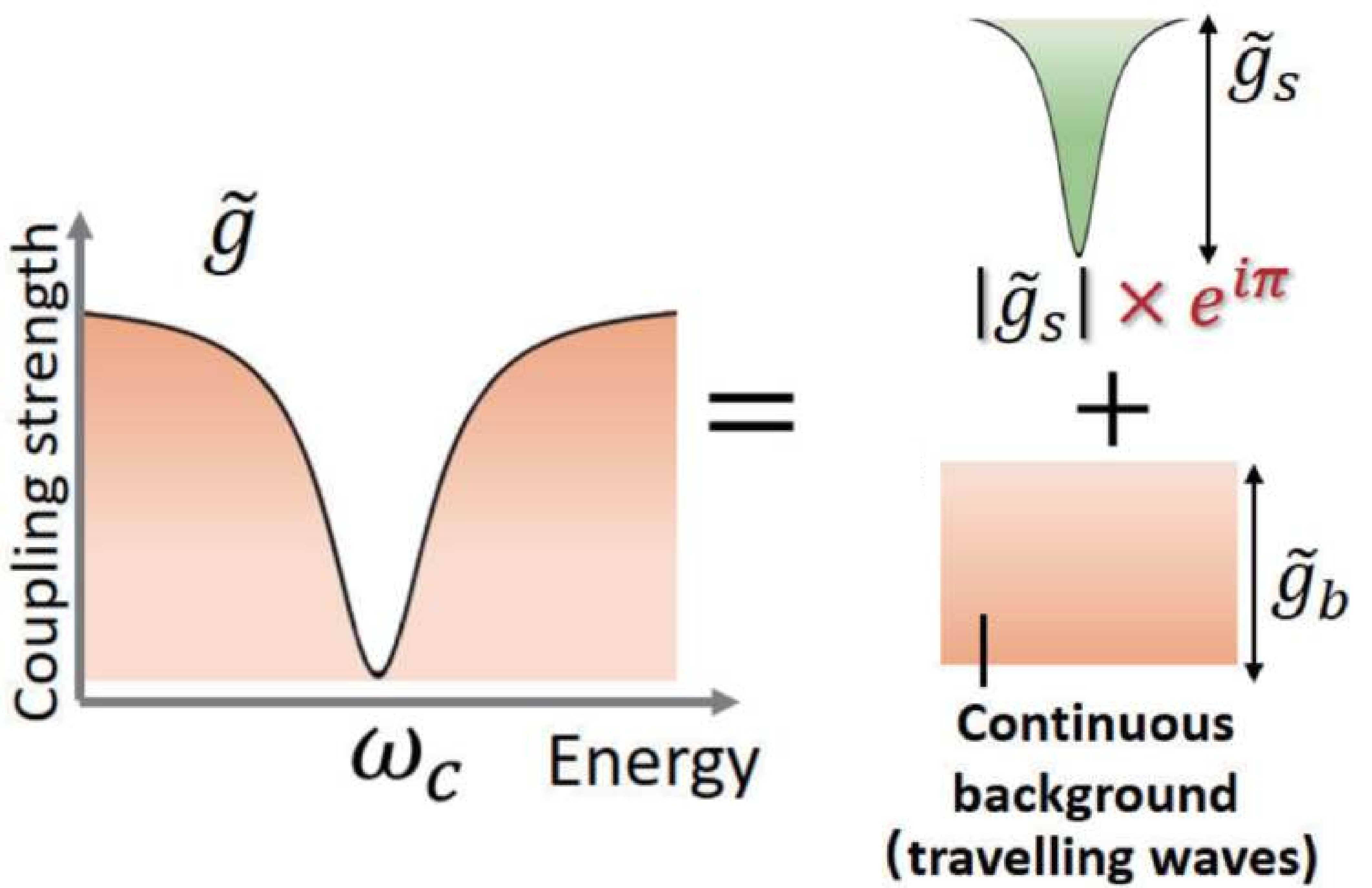} &  \begin{center}
			\begin{flushleft}
				${\small\bullet}$~Photon density of state (DOS)\\
			\end{flushleft}\begin{flushleft}
			${\small\bullet}$~Continuous background\\
		\end{flushleft}\begin{flushleft}
		${\small\bullet}$~Dissipative coupling\\
	\end{flushleft}\begin{flushleft}
	${\small\bullet}$~Level attraction
\end{flushleft}
		\end{center}\\ \\
		\begin{center}
		Yu \textit{et al.} \cite{Yu-19}\\Metelmann \textit{et al.} \cite{Clerk-15}
	\end{center}
	&
	\includegraphics[width=3.6cm]{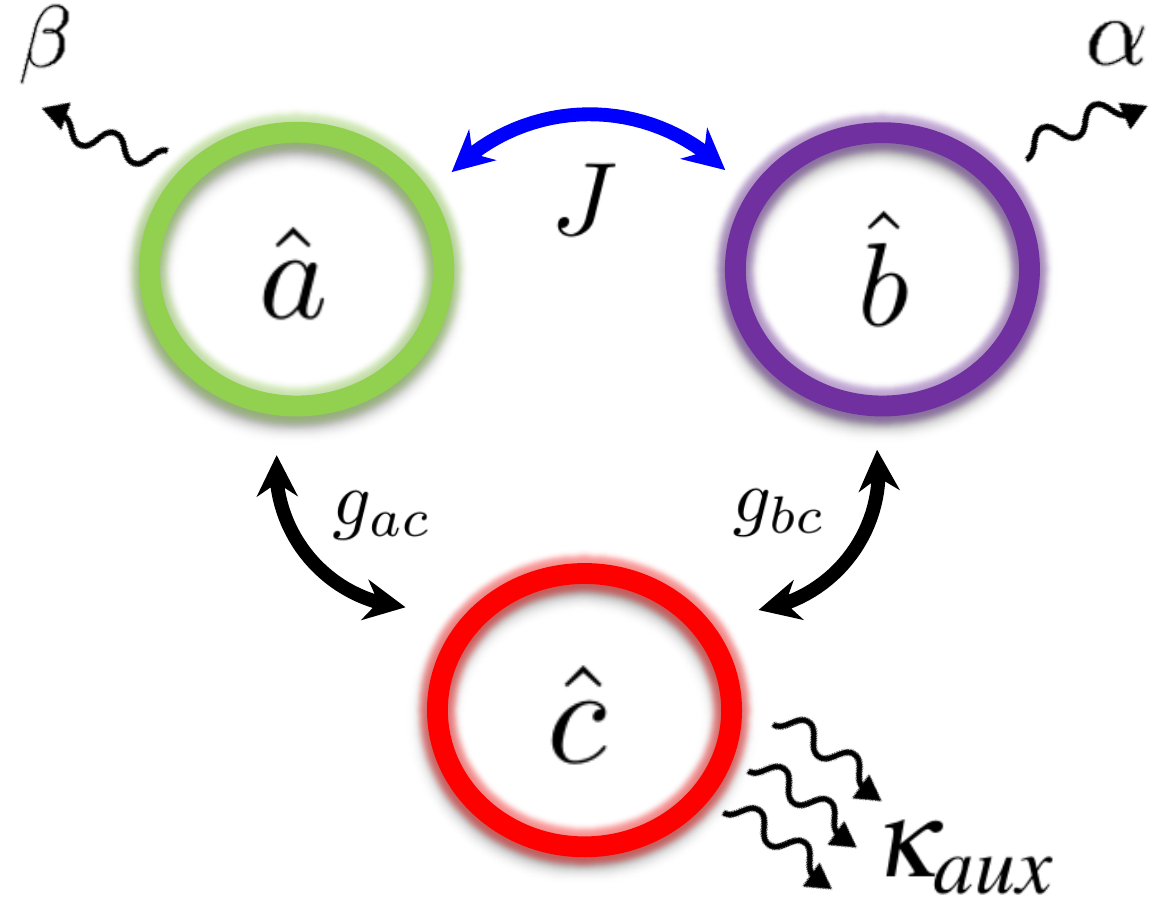} &  \begin{center}
		\begin{flushleft}
			${\small\bullet}$~Auxiliary cavity mode\\
		\end{flushleft}\begin{flushleft}
		${\small\bullet}$~Dissipative coupling
	\end{flushleft}
	\end{center}\\ \\
		\begin{center}
			Grigoryan \textit{et al.} \cite{Xia-18}
		\end{center}&
		\includegraphics[width=3.9cm]{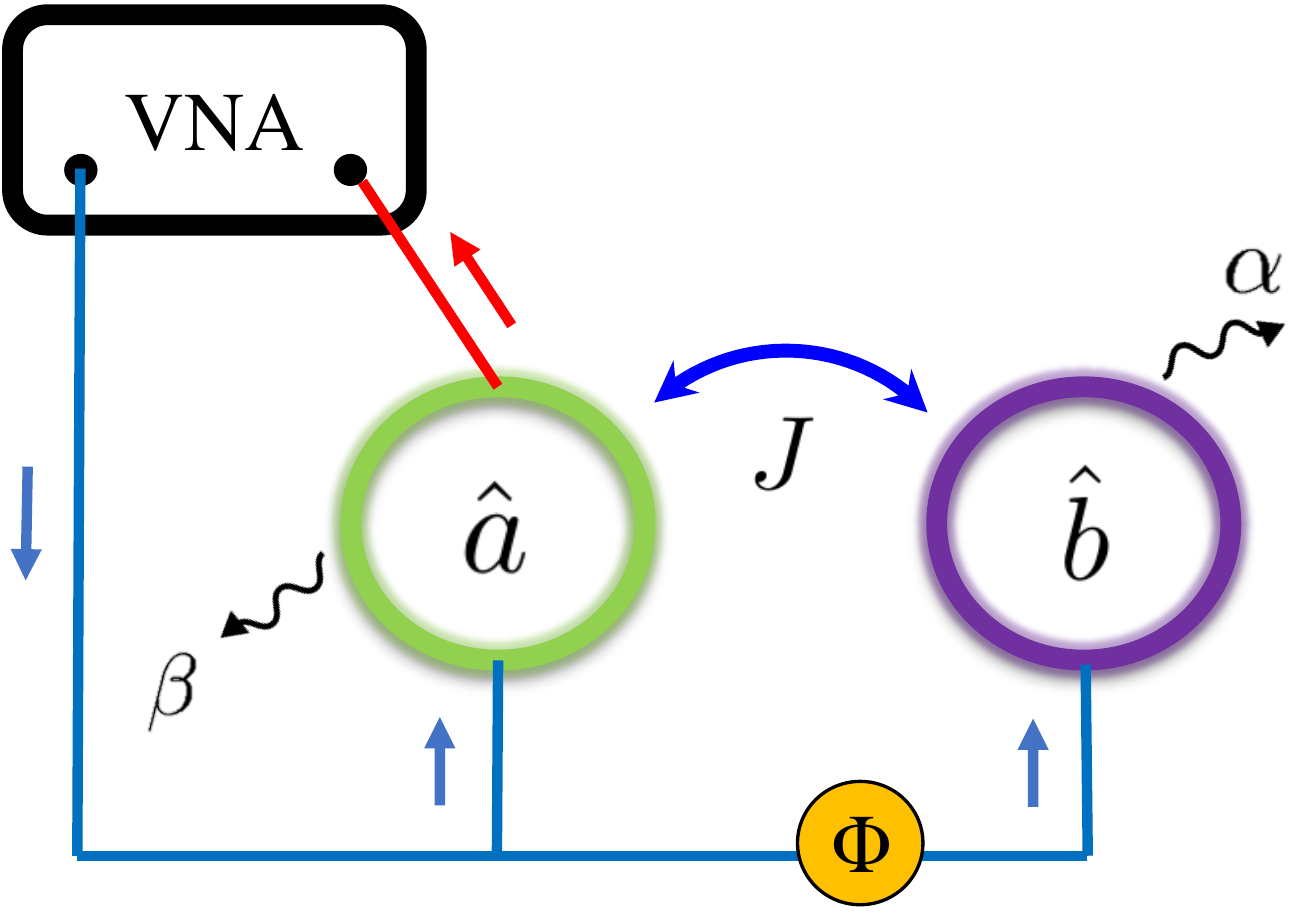} &  \begin{center}
			\begin{flushleft}
				${\small\bullet}$~Two-tone driven\\
			\end{flushleft}\begin{flushleft}
			${\small\bullet}$~Interference\\
		\end{flushleft}\begin{flushleft}
		${\small\bullet}$~Level attraction
	\end{flushleft}
		\end{center}\\
		\bottomrule
	\end{tabular}
	\label{Table2}
\end{table*}

\section{Comparison of different Theoretical pictures}\label{Theory}
In this section, we will introduce as detailed as possible the sequential results achieved by different research groups in the past few years. To provide some guidance to the relevant literatures and their main features, a summary of several key studies is provided in Table~\ref{Table2}. By following these studies, the underlying mechanism will be revealed. It will be found that dissipation is an essential factor for the formation of dissipative coupling. We also discuss the commonality and differences of traveling wave-type and cavity mode-type dissipative reservoirs in sustaining dissipative coupling.

\subsection{Electrodynamic picture: Cavity Lenz effect}\label{sec3A}

\begin{figure}[!b]
	\centering
	\includegraphics[width=0.46\textwidth]{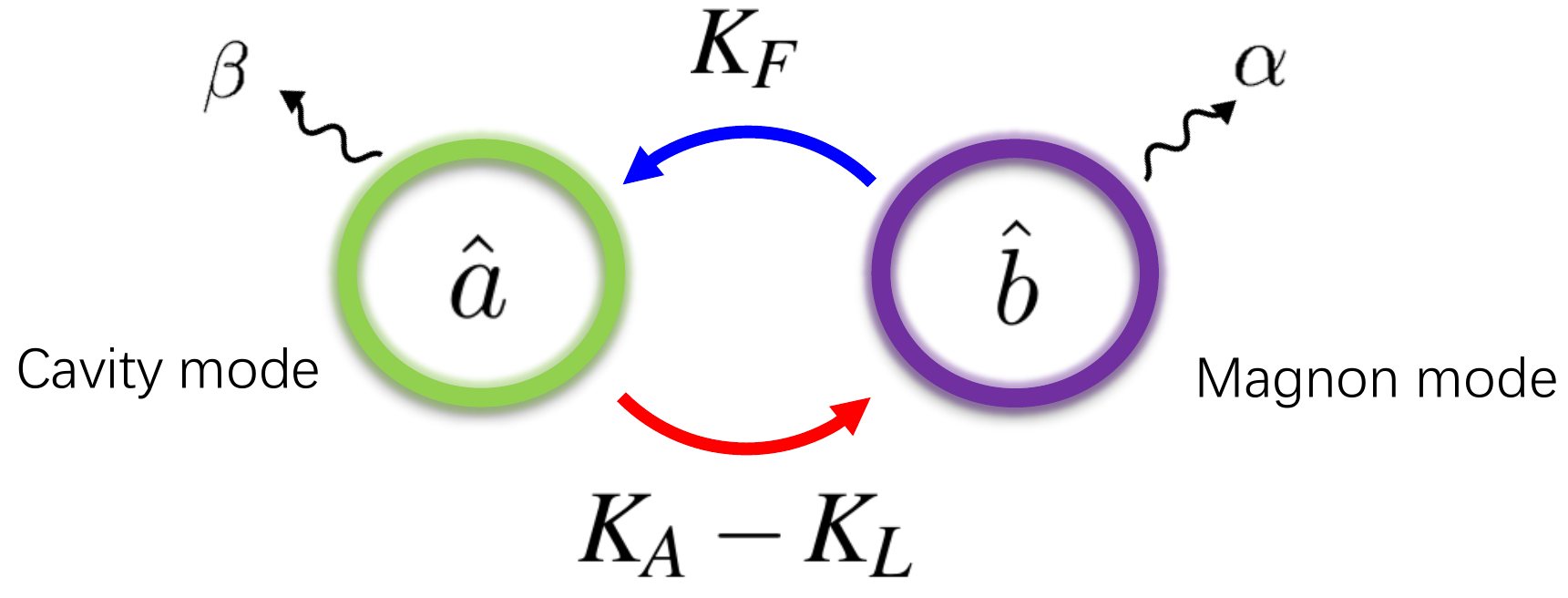}
	\caption{(color online). Schematic diagram of the phenomenlogical Amp\`{e}re effect ($K_{A}$), Faraday effect ($K_{F}$) and cavity Lenz effect ($K_{L}$) induced coupling between the cavity mode and magnon mode. The intrinsic damping rates of the cavity mode and magnon mode are $\beta$ and $\alpha$, respectively.  }
	\label{fig:2}
\end{figure}

Dissipative magnon-photon coupling was initially understood based on a phenomenological electrodynamic picture~\cite{Harder-18}. In that picture, Amp\`{e}re law describes that the microwave current $j$ produces an rf magnetic field, which drives the magnetization of the YIG sample via a driving torque. Faraday effect shows the influence of the dynamic magnetization $m$ on the rf current $j$. The dissipative coupling is modeled by the cavity Lenz effect, which describes the backaction from the induced rf current impeding the magnetization dynamics, instead of driving it. The equations of the cavity current $j$ and dynamic magnetization $m$ can be written as~\cite{Harder-18}:
\begin{equation}
\begin{pmatrix} \omega^2-\omega_c^2+i2\beta\omega_c\omega & i\omega^2K_F\\-i\omega_0(K_A-K_L) & \omega-\omega_m+i\alpha\omega \end{pmatrix} \begin{pmatrix} j \\ m \end{pmatrix} = 0,
\label{coupling}
\end{equation}
where $\omega_{0}=\gamma_{e}M_{0}$ is proportional to the saturation magnetization $M_{0}$, and $\gamma_{e}$ is the gyromagnetic ratio. This equation describes the coupling between the magnon mode at the frequency of $\omega_m$ and the cavity mode at $\omega_{c}$. The intrinsic damping rate of the magnon mode and cavity mode are $\alpha$ and $\beta$, respectively. The $K_{F}$ term stems from Faraday's law. The $K_{A}$ term comes from Amp\`{e}re law. The most interesting term is the $K_{L}$ term, which represents the cavity Lenz effect~\cite{Harder-18} and has the opposite sign with respect to the $K_{A}$. The competition between the $K_{L}$ and $K_{A}$ terms will lead to a net coherent or dissipative coupling. The coupled system can be equivalently modeled by a phenomenal non-Hermitian Hamiltonian:
\begin{equation}\label{lenz}
H = \hbar \omega_c \hat{a}^\dagger \hat{a} + \hbar \omega_m \hat{b}^\dagger \hat{b} + \hbar g \left(\hat{a}^\dagger \hat{b} + e^{i\Phi} \hat{b}^\dagger \hat{a}\right),
\end{equation}
where $a^\dagger$ ($a$) and $b^\dagger$ ($b$) are the creation (annihilation) operators for cavity photons and magnons, respectively. Here, the coupling phase $\Phi$ describes the competing effect of two forms of magnon-photon coupling, and $g$ is the net coupling strength. When $\Phi$ equals to 0 and $\pi$, the coupling term describes the coherent coupling and dissipative coupling, respectively.

The cavity Lenz effect provides a good phenomenological interpretation that captures the essential characteristics of the dissipative coupling. As defined in the Supplementary Materials of Ref.~\cite{Harder-18}, the negative $K_{L}$ term stems from the damping-like torque acting on the magnetization. It is caused by the Lenz effect that describes a backaction of the magnetization dynamics due to its coupling to the microwave current. This model enables an intuitive understanding of the dissipative magnon-photon coupling based on the classical picture. The effective non-Hermitian Hamiltonian [Eq.~(\ref{lenz})] describes very well the measured hybrid modes dispersion in both level repulsion and attraction regimes~\cite{Harder-18}. However, this model does not reveal the quantum mechanical origin of the dissipative coupling. As we will review below, when both the magnon and cavity modes couple to the same dissipative channel, their cooperative effect induces an effective dissipative coupling between the two modes, as captured by the damping-like torque term described by the cavity Lenz effect.

\subsection{Cooperative radiative damping picture}\label{sec3B}
\begin{figure}[!b]
	\centering
	\includegraphics[width=0.48\textwidth]{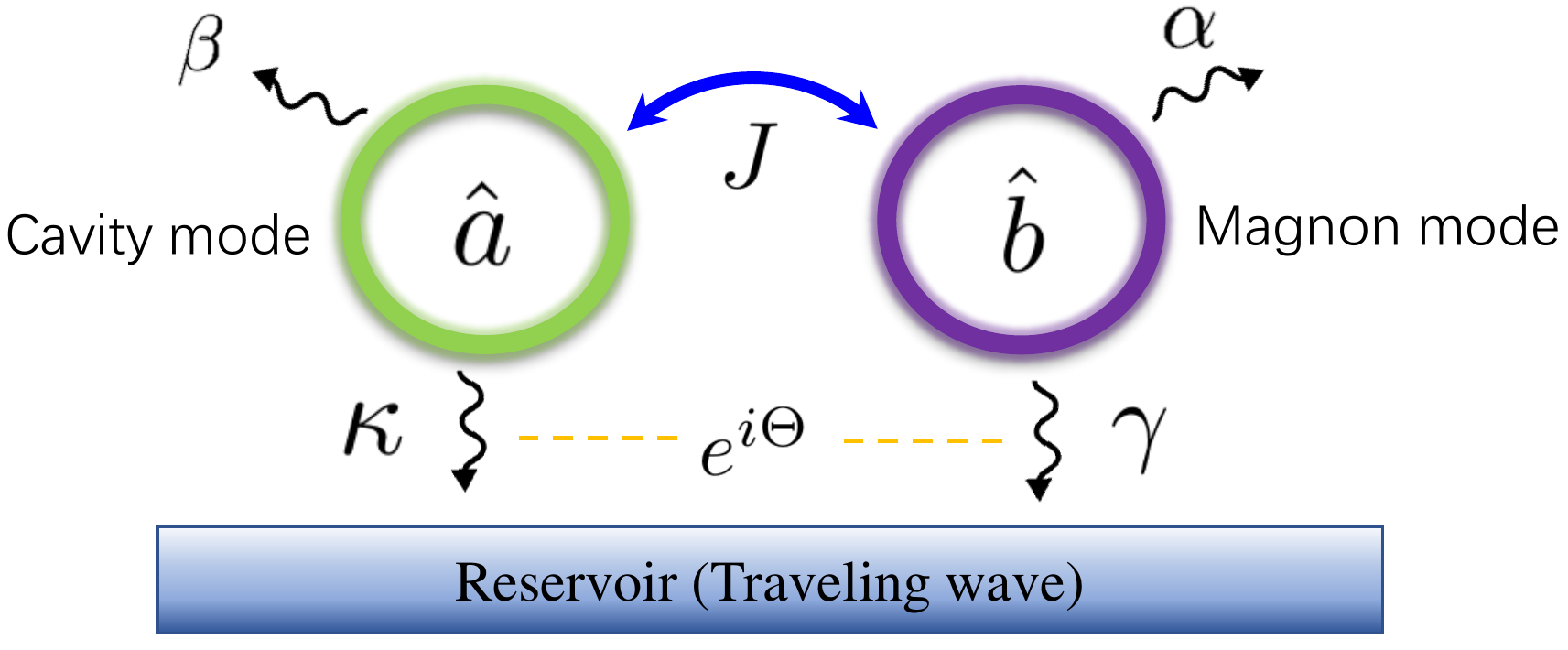}
	\caption{(color online). Schematic diagram of the traveling wave mediated dissipative coupling between the cavity mode and magnon mode. The cavity mode and magnon mode cooperatively damp to the traveling wave. The external damping rates of the two modes are $\kappa$ and $\gamma$, respectively. The dissipative coupling strength $\Gamma$ between these two modes has a simple relation of $\Gamma=\sqrt{\kappa\gamma}$. The external damping processes may have a relative phase $\Theta$. The cavity mode and magnon can also directly interact with each other with coherent coupling strength $J$. }
\label{fig:3}
\end{figure}

In light of the nonreciprocal microwave transmission in cavity magnonics experimentally observed by Wang $et~al.$ \cite{Wang-19}, a cooperative radiative damping picture, has been established. It reveals the origin of the dissipative coupling. This picture roots in the theory of reservoir engineering developed by Metelmann \textit{et al.} \cite{Clerk-15}. The schematic diagram is shown in Fig.~\ref{fig:3}. The magnon mode and cavity mode have intrinsic damping rates of $\alpha$ and $\beta$, respectively. Moreover, they both interact with the same reservoir (i.e., the traveling wave), which causes the external dampings of these two modes. The external damping rates are $\gamma$ and $\kappa$ for the magnon and cavity mode, respectively. Dissipative coupling between these two modes is sustained by their cooperative external dissipations. Now we begin with the von-Neumann-Lindblad master equation to see how cooperative external dampings are introduced and how it contributes to the dissipative coupling between the concerned modes. The equation is written as:
\begin{equation}\label{s1}
\frac{d}{dt}\hat{\rho}=-\frac{i}{\hbar}[\hat{H}_{coh},\hat{\rho}]+\hat{L}\hat{\rho},
\end{equation}
where the Hamiltonian $\hat{H}_{coh}$ describes the conserved energy of the relevant modes and the coherent coupling mechanism between them. The Lindblad operator $\hat{L}$ describes the dissipation consisting of the intrinsic damping rates of the coupled system and the external damping rates of the two modes due to their interaction with the traveling wave. Under the rotating wave approximation, the Hamiltonian $\hat{H}_{coh}$ has the form:
\begin{equation}\label{s2}
\hat{H}_{coh}/\hbar=\omega_{c}\hat{a}^{\dagger}\hat{a}+\omega_{m}\hat{b}^{\dagger}\hat{b}+J(\hat{a}^{\dagger}\hat{b}+\hat{a}\hat{b}^{\dagger}),
\end{equation}
where $\hat{a} (\hat{a}^\dag)$ and $\hat{b} (\hat{b}^\dag)$ are the annihilation (creation) operators of photon and magnon, respectively. $\omega_{c}$ and $\omega_{m}$ represent the frequencies of the uncoupled modes. $J$ is the coupling rate that describes the coherent magnon-photon coupling.

The standard dissipative superoperator $L[\hat{o}]$ is defined as:
\begin{equation}\label{s4}
L[\hat{o}]\hat{\rho}=\hat{o}\hat{\rho} \hat{o}^{\dag}-\frac{1}{2}\hat{o}^{\dag}\hat{o}\hat{\rho}-\frac{1}{2}\hat{\rho} \hat{o}^{\dag}\hat{o},
\end{equation}
where $\hat{o}$ is the jump operator, for different dynamical processes the jump operators have different forms. For the cooperative dissipations of the cavity and magnon mode, the jump operator is a linear superposition of the annihilation operators, that is
\begin{equation}\label{s5}
\nu \hat{a}+ue^{i\Theta} \hat{b}.
\end{equation}
Here, the coefficients $\nu$ and $u$ characterize the individual couplings of the cavity and magnon modes with the traveling wave and $e^{i\Theta}$ describes the relative phase between the couplings. If the spatial separation of the correlated modes is relatively small compared with the wavelength of the resonance frequency, then $\Theta\approx0$. The jump operator has the general form~\cite{Clerk-15,Wang-19}:
\begin{equation}\label{s5}
\hat{o}\rightarrow \nu \hat{a}+u \hat{b},
\end{equation}
which leads to the dissipative magnon-photon coupling. By adopting this cooperative jump operator, the general master equation can be written as:
\begin{equation}\label{s6}
\frac{d}{dt}\hat{\rho}=-\frac{i}{\hbar}[\hat{H}_{coh},\hat{\rho}]+\tau L[\hat{o}]\hat{\rho}+\beta L[\hat{a}]\hat{\rho}+\alpha L[\hat{b}]\hat{\rho},
\end{equation}
where the second term on the right-hand side describes the external dissipations of the two modes that are proportional to the rate $\tau$. The third and fourth terms represent the intrinsic dissipations of the cavity and magnon modes, with the rates $\beta$ and $\alpha$, respectively. The external damping rates induced by the traveling wave for the cavity and magnon modes are $\tau\cdot\nu^{2}=\kappa$ and $\tau\cdot u^{2}=\gamma$, respectively. Furthermore, from the second term, we can get the cooperative damping rate of the cavity and magnon modes is $\tau\cdot\nu u=\sqrt{\kappa\gamma}$.

The cavity mode and magnon mode can directly interact with each other with coherent coupling strength $J$. The equations of motion for the cavity mode $\hat{a}$ and magnon mode $\hat{b}$ can be derived as:
\begin{eqnarray}
\frac{d}{dt}\hat{a}&=&-i\omega_{c}\hat{a}-(\beta+\kappa)\hat{a}-\big(iJ+\sqrt{\kappa\gamma}\big)\hat{b}, \nonumber \\
\frac{d}{dt}\hat{b}&=&-i\omega_{m}\hat{b}-(\alpha+\gamma)\hat{b}-\big(iJ+\sqrt{\kappa\gamma}\big)\hat{a},
\end{eqnarray}
it is very clear that the total coupling between the cavity mode and magnon mode is measured by the coefficient $iJ+\sqrt{\kappa\gamma}$. We can find the dissipative coupling strength $\Gamma$ has a simple and elegant relation with external damping rates, which is $\Gamma=\sqrt{\kappa\gamma}$. Then the effective coupling strength between the cavity mode and magnon mode is $J-i\Gamma$.
Hence, we can write the effective non-Hermitian Hamiltonian of the coupled system as:
\begin{equation}\label{Hamiltonian2}
\hat{H}/\hbar=\widetilde{\omega}_{c}\hat{a}^{\dag}\hat{a}+\widetilde{\omega}_{m}\hat{b}^{\dag}\hat{b}+(J-i\Gamma)( \hat{a}^{\dag}\hat{b}+\hat{a}\hat{b}^{\dag})\\
\end{equation}
where $\widetilde{\omega}_{c}$ and $\widetilde{\omega}_{m}$ represent the complex eigenvalues of the uncoupled cavity and magnon modes, which are defined as $\widetilde{\omega}_{c}=\omega_{c}-i\beta$ and $\widetilde{\omega}_{m}=\omega_{m}-i\alpha$. Further, we have mentioned above that the cooperative couplings of the two modes to the traveling wave may have a relative phase. It can be described by the phase term in the jump operator, and this phase term finally attributed to the relative phase between the coherent and dissipative couplings.

It should be noted that up to now, most experimental results of the dissipative coupling in cavity magnonics can be attributed to this mechanism~\cite{Harder-18,Bhoi-19,Ying-19,Rao-19,Wang-19,Rao-19-2,Yao-19}.

\subsection{Photon density of state picture}\label{sec3C}
The cooperative radiative damping picture described in the previous section shows that the traveling wave is the key ingredient that induces the magnon-photon dissipative coupling. That is the common feature for all cavities summarized in the first four rows of the Table~\ref{Table1} where LA is observed. All of them support the propagation of the traveling wave and formation of the standing cavity mode. Among them, the 1D waveguide-cavity setup~\cite{Harder-18,Yao-19,Yao-19-2} can adjust the weights of the traveling wave and standing waves by tuning the global geometry via rotating the relative angle $\theta$ between the two waveguide transitions. To capture such a wave controllability, Yao \textit{et al.} \cite{Yao-19,Yao-19-2} developed a theory for calculating the photon density of state, by considering all the photon modes and carefully distinguishing the standing and traveling wave parts. This theoretical picture provides a good understanding of the difference between the effect of traveling and standing waves on light-matter interaction.

The superposition of these two contributions can be monitored by the radiative linewidth of the hybridized mode. At resonance, the damping rates of the coherently coupled subsystems will attract with each other, and the linewidths of the dissipatively coupled modes repulse with each other, as shown in Figs.~\ref{fig:1}(e) and 1(f), respectively. In the case of magnon mode linewidth smaller than the cavity mode linewidth, the standing wave photon states contribute to the coherent coupling and linewidth broadening of the magnon mode, and the continuous wave photon states contribute to the dissipative coupling and linewidth shrink of the magnon mode.

Yao \textit{et al.} \cite{Yao-19,Yao-19-2} show that quantitatively, the amount of the radiative linewidth changing of the magnon-like hybridized mode can be mapped to the density of photon states. It provides another way for verifying the influence of the traveling wave on the effect of magnon-photon coupling.

\subsection{Dissipative coupling mediated by a damped auxiliary cavity mode}\label{sec3D}
While the traveling wave is the key ingredient of the open cavity magnonic systems that leads to the observation of LA, it is not the only way for inducing dissipative coupling. A few theories have proposed an alternative mechanism for realizing dissipative couplings by utilizing an auxiliary cavity mode~\cite{Clerk-15,Yu-19,Du-19}, which is very different from the mechanism of the traveling wave. This auxiliary mode, with a large intrinsic damping rate, can be seen as a common reservoir for inducing indirect coupling between two principle modes. The induced indirect dissipative coupling strength depends on the damping rate of the auxiliary mode, the frequency detunings, and the direct coupling strengths between the principle modes and the auxiliary mode. Simply speaking, it is not very easy for a damped auxiliary cavity mode to support a purely large dissipative coupling strength.
\begin{figure}[!b]
	\centering
	\includegraphics[width=0.46\textwidth]{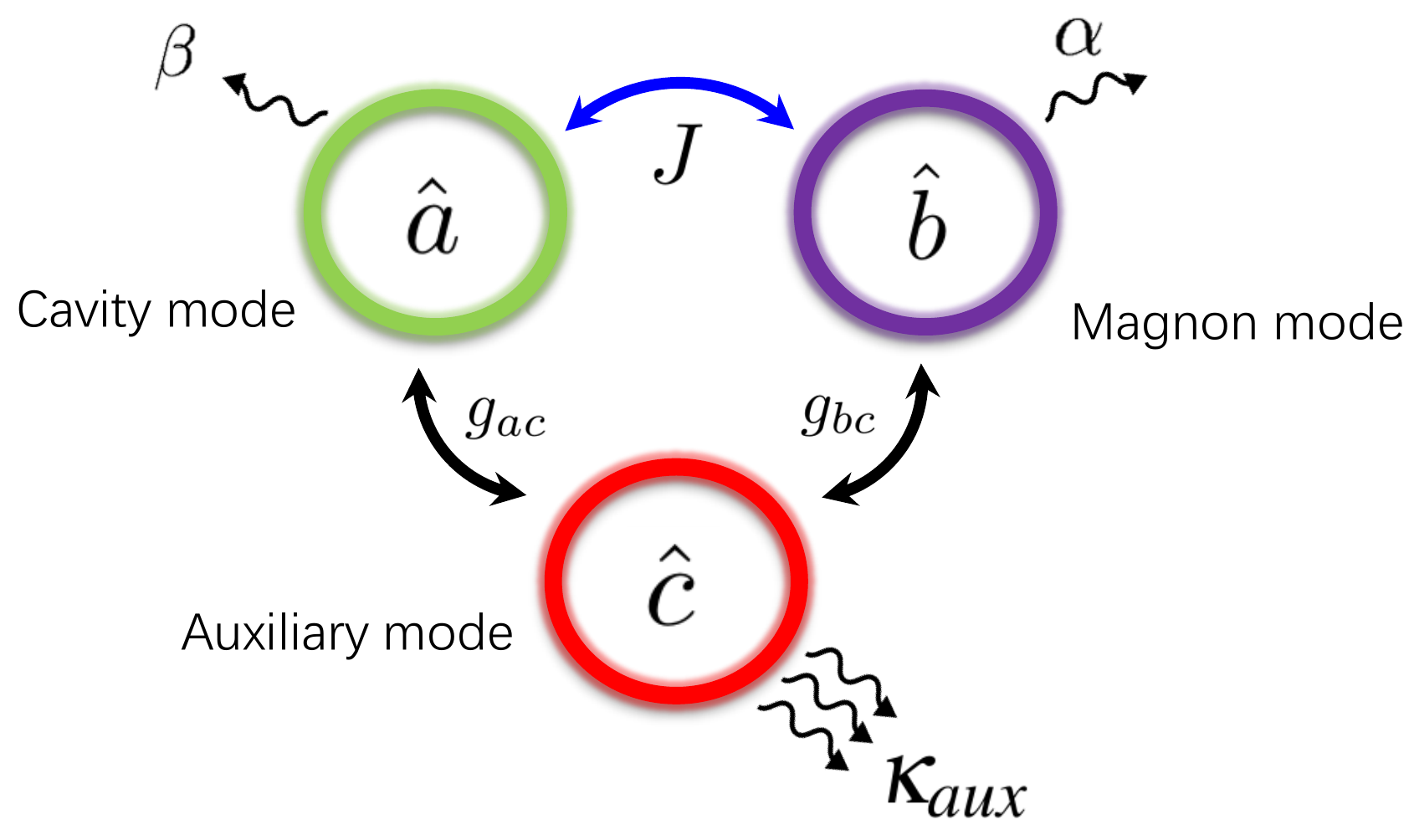}
	\caption{(color online). Schematic diagram of the auxiliary damped mode mediated coupling between the cavity mode and magnon mode. The intrinsic damping rates of the cavity mode and magnon mode are $\beta$ and $\alpha$, respectively. The damping rate of the auxiliary mode is $\kappa_{aux}$. The coherent coupling strength between the cavity (magnon) mode and auxiliary mode is $g_{ac(bc)}$. The cavity mode and magnon mode can also directly interact with each other with coupling strength $J$.  }
	\label{fig:4}
\end{figure}

Taking the auxiliary mode proposed by Yu \textit{et al.}~\cite{Yu-19} as an example, Fig.~\ref{fig:4} schematically shows the sustained dissipative coupling between the cavity mode and magnon mode. The cavity mode and magnon mode have intrinsic damping rates $\beta$ and $\alpha$, respectively. Both of them are coherently coupled with the auxiliary mode with the coupling strength $g_{ac}$ and $g_{bc}$, respectively. The intrinsic damping rate of the auxiliary mode is $\kappa_{aux}$. The cavity mode and magnon mode can also interact coherently with each other, with the coupling strength $J$. In the dispersive limit and assuming the cavity mode is resonant with magnon mode and $J=0$, the effective coupling sustained by the auxiliary mode is:
\begin{equation}\label{s5}
g_{eff}=-i\frac{g_{ac}g_{bc}}{\kappa_{aux}-i\Delta},
\end{equation}
where $\Delta$ is the frequency detuning between the auxiliary mode and cavity mode (magnon mode). Eq.~(\ref{s5}) shows that a dissipative coupling (with an imaginary coupling rate) can be induced by the auxiliary mode.

\begin{figure}[!b]
	\centering
	\includegraphics[width=0.48\textwidth]{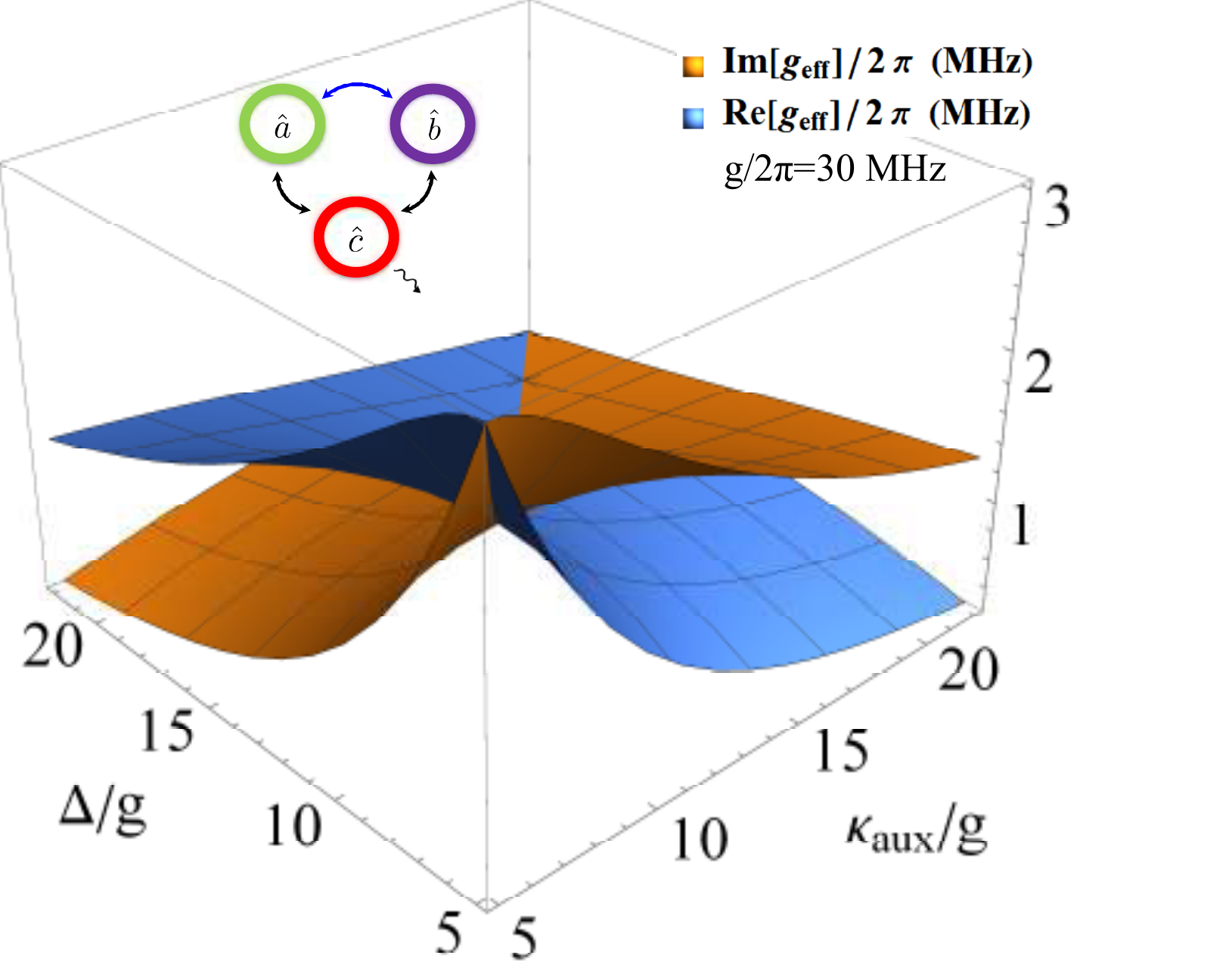}
	\caption{(color online). The imaginary and real parts of the effective coupling strength ($g_{eff}$) between the cavity mode and magnon mode sustained by the auxiliary mode as a function of the detuning $\Delta$ between the cavity (magnon) mode and the auxiliary mode and the damping rate $\kappa_{aux}$ of the auxiliary mode. For simplicity, we assume the cavity mode and magnon mode are in resonance. The coherent coupling strength between the cavity (magnon) mode and the auxiliary mode $g_{ac(bc)}$ is chosen to equal to $g/2\pi=30$ MHz.}
	\label{fig:5}
\end{figure}

We have mentioned that while both a traveling wave and an auxiliary mode can induce the dissipative coupling, the induced coupling strength is very different. In the case of a traveling wave as described in the section \ref{sec3B}, if the spatial separation of the two coupled modes is relatively small compared with the wavelength of the resonance frequency, the induced coupling is purely dissipative, with the imaginary coupling rate given by $i\Gamma=i\sqrt{\kappa\gamma}$ which increases with the external damping rates.  On the contrary, for the dissipative coupling induced by the auxiliary mode, the induced coupling is generally a complex number as given by Eq.~(\ref{s5}), which involves both coherent and dissipative couplings. The nearly pure dissipative coupling can only be realized in the case of dispersive limit but with a small detuning [$g_{ac (bc)} \ll \Delta \ll \kappa_{aux}$], where the coupling rate $g_{eff}=-i g_{ac}g_{bc}/\kappa_{aux}$ decreases when the damping rate of the auxiliary mode increases.

To visualize the complex coupling strength induced by the auxiliary mode, we assume $g_{ac}/2\pi=g_{bc}/2\pi=g/2\pi=30$ MHz and plot the real part (coherent coupling strength) and imaginary part (dissipative coupling strength) of $g_{eff}/2\pi$ in Fig.~\ref{fig:5} as a function of $\Delta/g$ and $\kappa_{aux}/g$. Here, we restrict the case in the dispersive limit with a low damping auxiliary mode, where the detuning $\Delta$ and damping rate $\kappa_{aux}$ are at least five times larger than the coupling strength $g$. Clearly, the largest dissipative coupling strength is obtained when both of the detuning $\Delta$ and damping $\kappa$ are small. If a pure dissipative coupling is desired, the damping rate of the auxiliary mode should be much larger than the detuning, but that will also lead to a relatively small dissipative coupling strength.

The difference between the traveling wave-type reservoir and damped auxiliary cavity mode-type reservoir can also be viewed through the picture of the local photon density of states (LDOS) of the vibrational electromagnetic environment~\cite{Yao-19-2}, which is described in the section \ref{sec3C}. The traveling wave can be seen as a continuous photon state while the damped auxiliary cavity mode only supports the photon modes near its eigenfrequencies. This leads to the difference in the induced dissipative coupling strength.


\subsection{Level attraction induced by the two-tone driven scheme}\label{sec3E}
While LA is useful for studying the dissipative coupling, the effect of LA itself is a ubiquitous phenomenon not limited by the dissipative coupling. For example, LA can also emerge in coherently coupled systems via diverse mechanisms such as two-tone excitation, parametric driving, and coherent coupling between normal and inverted oscillators. In this section, we discuss the two-tone driven scheme that can induce LA.

First, recalling the experimental configuration of two-tone driven coupled cavity photon-magnon system~\cite{Boventer-19,Boventer-19-2}, in addition to the magnetic component of the cavity microwave field, a local microwave magnetic field is applied to excite the magnons as depicted in the last row of Table~\ref{Table2}. The two driven tones are phase-correlated. The anti-crossing gap between the hybridized modes can be controlled by the relative phase $\Phi$ between the two driving forces and their intensity ratio of $\delta$. The effective coupling strength between the photon mode and magnon mode is $(1+\delta e^{i\Phi})K$, where $K$ is the coupling strength when only the cavity microwave field is applied.

In the $\Phi=0$ case, the anti-crossing gap is enhanced due to the local driving field. In contrast, when $\Phi=\pi$ and $\delta>1$, the gap is reduced, leading to the synchronization of the coupled modes~\cite{Xia-18}. In this case, the two driving fields exert opposite torque on the magnetization of the YIG and cancel with each other.

Although the explanations given in this two-tone driven scheme was phenomenological without considering the underlying mechanism, they may also be closely related to the dissipation channel induced effects. In two-tone driven experimental devices, due to the presence of an additional RF loop introduced into the cavity, which interfaces both magnons as well as the cavity photons. The signal from this loop has the nature of the traveling wave, which may support the emergence of dissipative coupling. Hence, it requires further detailed research and clarification of the mechanism of the “two-tone driven” scheme. Nevertheless, it should be cautious that observing the LA phenomena in the cavity magnonics system is not sufficient for verifying dissipative couplings.


\section{Outlook}\label{sec4}
In this section, we discuss some interesting directions for future studying and utilizing the dissipative coupling mechanism.

\subsection{Interplay between the coherent and dissipative couplings}
Both the coherent and dissipative couplings are ubiquitous in natural systems, and a hybrid system governed by both the coherent and dissipative couplings is of particular interest for future investigation.

First of all, how would we monitor and analyze the interplay between the coherent and dissipative couplings? Transmission spectroscopy may provide the most convenient method for such a purpose. Following the Eqs.~(\ref{eq:input-output}) and (\ref{Hamiltonian2}), the transmission coefficient of a cavity photon-magnon system is~\cite{Ying-19,Wang-19}:
\begin{equation}\label{s13}
S_{Trs}=1+\frac{\kappa }{i(\omega-\omega_{c})-(\kappa+\beta)+\frac{-(iJ+\Gamma)^{2}}{i(\omega-\omega_{\rm{m}})-(\alpha+\gamma)}},
\end{equation}
where the coupling strength ($J-i\Gamma$) is complex due to the coherent and dissipative couplings.

\begin{figure}[!t]
	\centering
	\includegraphics[width=0.48\textwidth]{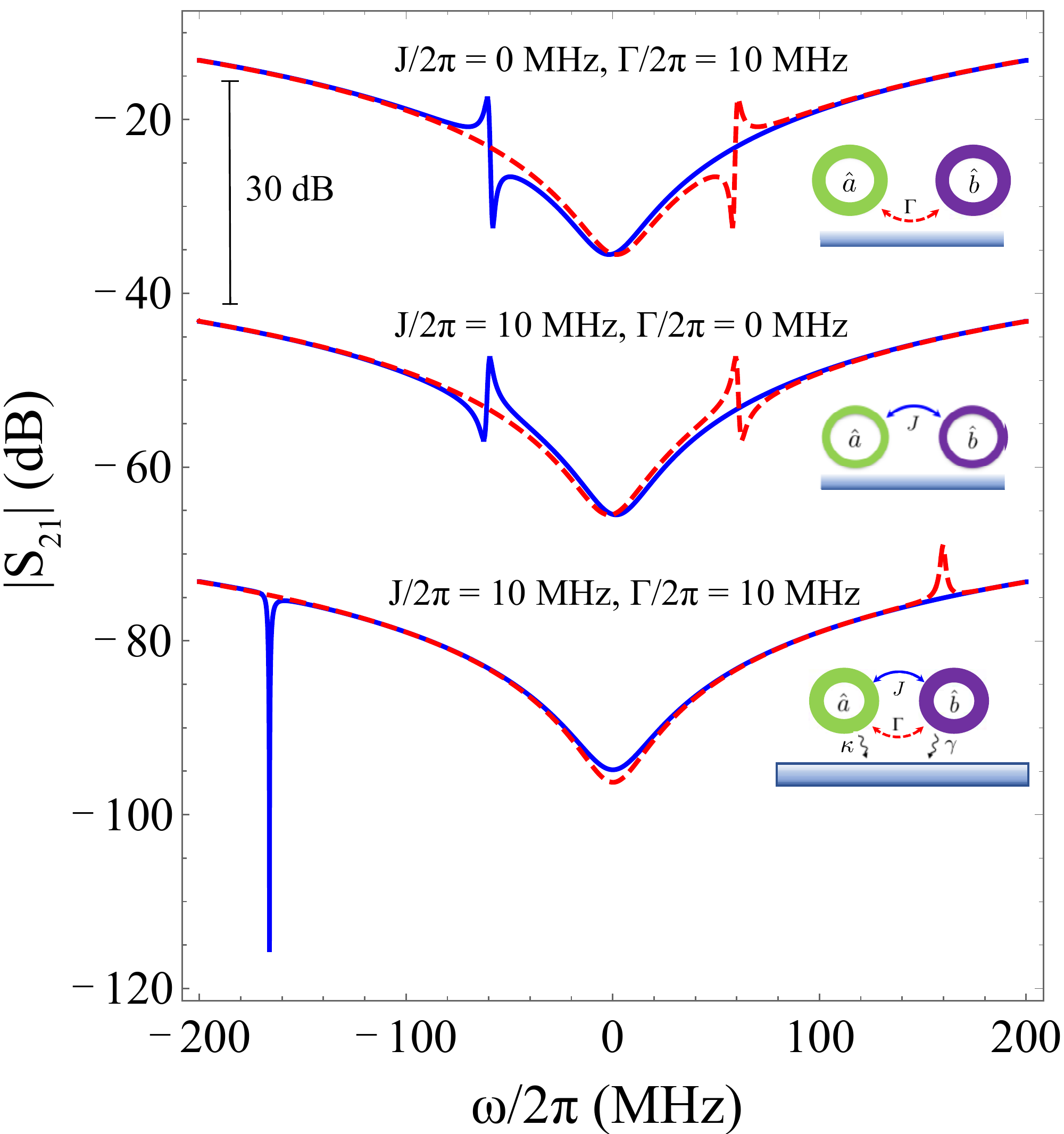}
	\caption{(color online). Transmission spectra of the coupled system.  The intrinsic damping rates of the cavity mode and magnon mode are 15 MHz and 1.1 MHz, respectively. The external damping rate of the cavity mode is 880 MHz. Top panel: Coherent coupling strength $J/2\pi$= 0 MHz, and dissipative coupling strength $\Gamma/2\pi$=10 MHz. Middle panel: $J/2\pi$= 10 MHz, $\Gamma/2\pi$=0 MHz. Bottom panel: $J/2\pi$= 10 MHz, $\Gamma/2\pi$=10 MHz. The curves are plotted with 30 dB vertical offset for clarity.}
	\label{fig:6}
\end{figure}

A characteristic feature of such a complexly coupled system is that the transmission spectra are asymmetric with respect to the modes frequency detuning. We fix the cavity mode frequency as $\omega_{c}$ and detune the frequency of the magnon mode $\omega_{m}$, the transmission spectra of the coupled system at positive and negative field detunings ($\Delta_{m}=\omega_{m}-\omega_{c}$) are plotted in Fig.~\ref{fig:6} (bottom panel). The coherent coupling strength $J/2\pi$ and dissipative coupling strength $\Gamma/2\pi$ are both set at 10 MHz. Other damping parameters are chosen following the experimental work done recently~\cite{Wang-19}. The solid blue and dashed red curves are transmission spectra correspond to the field detunings $\Delta_{m}/2\pi$ equal to -160 MHz and 160 MHz, respectively. They are strongly asymmetric with respect to the center frequency. For comparison, we plot the transmission spectra of the coupled system when the coupling is purely coherent or dissipative in Fig.~\ref{fig:6} (top and middle panels). In both cases, the spectra are symmetric with respect to the detuning.

Utilizing such a characteristic spectral feature for monitoring and quantifying the complex coupling strength, various methods based on position and field tuning can be developed for controlling the interplay between the coherent and dissipative couplings. Stems from the interference between the coherent and dissipative couplings,  the coupled system exhibits nonreciprocal response~\cite{Clerk-15,Fang-17,Xu-19,Wang-19}, which can be utilized for developing novel isolators, circulators, and even directional amplifies. This scheme of isolation is different from the conventional nonreciprocal devices utilizing either Faraday rotation or ferromagnetic resonance~\cite{Hogan-53,Rowen-53,Adam-2002,Camley-09}, where large ferrites and complex port design (involving resistive sheets, quarter-wave plates, etc.) are needed. These components make the device bulky for integration. In our new scheme, as long as the interference between the coherent coupling and dissipative coupling can be achieved, parameters such as the volume and shape of the ferrites are not crucial factors limiting the isolation ratio~\cite{Wang-19}.

Furthermore, also due to such interference effect, ultrasharp resonance appears in the transmission spectrum, as shown in the bottom panel of Fig.~\ref{fig:6} (Solid blue curve). This feature can be exploited for developing new sensing and switching technologies.

\subsection{Dissipation engineering in quantum systems}

\begin{figure}[!b]
	\centering
	\includegraphics[width=0.48\textwidth]{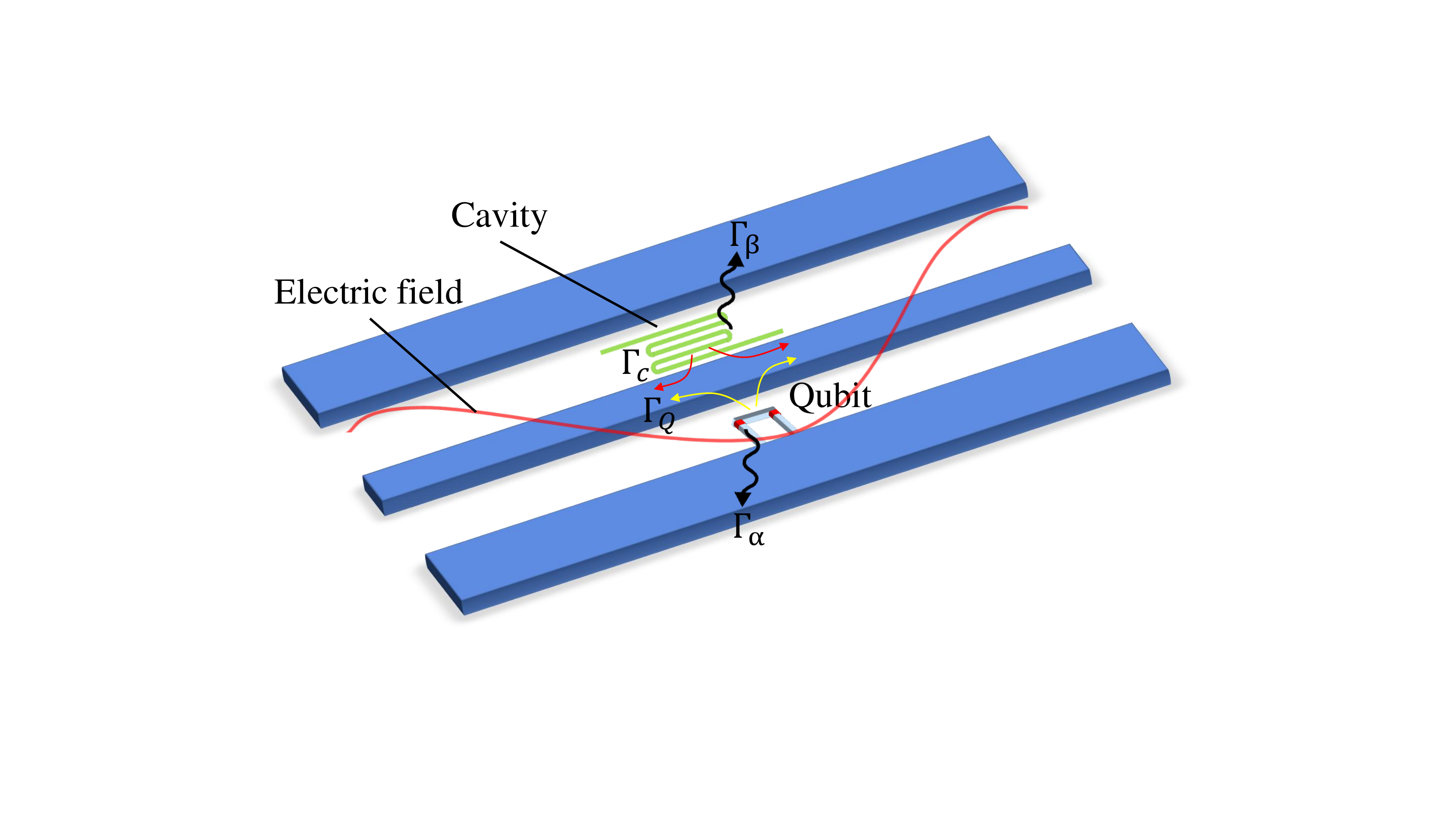}
	\caption{(color online). Schematic diagram of a waveguide QED system consists of a strip resonator (cavity) and a qubit that cooperatively coupled with a common waveguide. The rates of the cooperative damping to the waveguide are $\Gamma_{C}$ and $\Gamma_{Q}$ for the cavity mode and qubit, respectively.  }
	\label{fig:7}
\end{figure}

\begin{figure}[!t]
	\centering
	\includegraphics[width=0.48\textwidth]{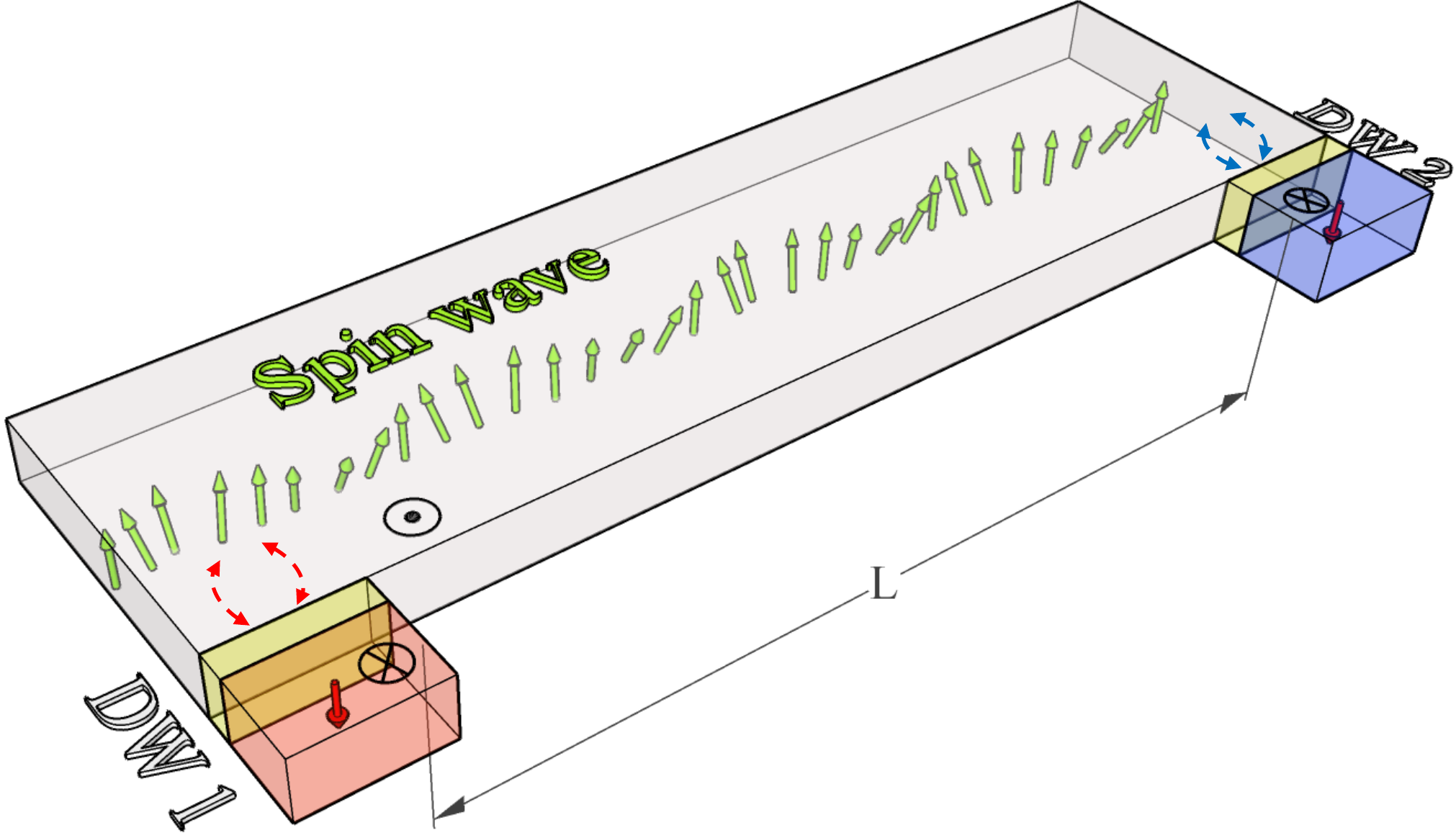}
	\caption{(color online). Schematic diagram of the propagating spin wave mediated remote coupling between two domain walls. }
	\label{fig:8}
\end{figure}

The physics of dissipative coupling discussed in this article is not limited in the cavity magnonics system, since dissipation is ubiquitous in quantum systems. For example, in the circuit-QED system~\cite{Devoret-13,You-11}, the superconducting qubit has been successfully used for quantum information processing. Such a quantum device is inevitably coupled to the environment via dissipation processes.  For decades, the coherent coupling has been studied and utilized in the circuit-QED system. In light of the physics of dissipative coupling revealed in the cavity magnonic systems, one may explore the cooperative damping mechanism in hybrid quantum systems involving qubits. As an example, Fig.~\ref{fig:7} shows a strip resonator (cavity mode) and a superconducting qubit coupled to the waveguide, with the radiation damping rates of $\Gamma_{C}$ and $\Gamma_{Q}$, respectively. It would be exciting to explore the dissipative coupling between the qubit and the strip resonator, which might be induced by their cooperative dampings. Following the same physical principle, one may also replace the strip resonator with other oscillators such as quantum dots or mechanical nano-oscillators. The introducing of dissipative coupling into such hybrid quantum systems may open new avenues for quantum information processing by utilizing dissipation and reservoir engineering\cite{Clerk-15}. 

In atomic physics and quantum optics, dissipative coupling discussed here has also been proposed and understood as the radiative coupling between dipoles where several dipoles can interact with each other through coupling to a common bath~\cite{Agarwal-book}. On the base of this framework, the dissipative coupling may also be useful in the implementation of quantum entanglement~\cite{Reiter-16,Kastoryano-11,Piani,Agarwal-86}.

\subsection{Harnessing dissipative coupling for cavity spintronics}
The development of spintronics demands new methods and techniques for generating, detecting, and manipulating spin currents. Via the magnon-photon coupling, cavity techniques, which have transformed many fields such as atomic physics and optics, have been introduced to the field of spintronics. For example, spin pumping~\cite{Silsbee,Bret-03,Tserkovnyak-05,Harder-PR} is an efficient method for generating spin currents via magnon excitations. Combining spin pumping with microwave cavity technique, experiments have detected spin currents generated by cavity magnon polaritons (CMP), which are hybrid excitations of magnon and cavity photon modes \cite{Hu-15,Maier-16}. Utilizing the non-local hybrid nature of CMP, coherent magnon-photon coupling becomes a useful method for distant manipulation and switch of spin currents \cite{Hu-17}, which is a striking example of potential capabilities of cavity spintronics.

Dissipative magnon-photon coupling may bring even more capabilities. For example, such a coupling leads to level attraction with a coalescence of hybridized magnon-photon modes, which is distinctly different from level repulsion with mode anticrossing caused by coherent magnon-photon coupling. This feature may be explored in devices where an array of magnetic tunnel junctions (MTJs) are cooperatively coupled to the same dissipative environment, to create synchronized oscillations. MTJs work as spin-torque nano-oscillators for generating continuous-wave microwave signals. The outstanding challenge is to find ways to synchronize them to enhance the output power.  Through the dissipative channel mediated synchronization by harnessing the dissipative magnon-photon coupling, a macroscopic ensemble of many nano-oscillators in the cavity spintronics device could become a powerful microwave source.

In a multi-domain spintronics device, the dissipative coupling mechanism is also implementable. As schematically shown in Fig.~\ref{fig:8}, we propose two spatially separated domain walls collectively couple to a dissipative spin wave bath. The propagating spin wave can construct and transfer coherence between the dissipative processes of these two domain walls; in other words, the spin wave serves as a dissipative channel for these two domain walls simultaneously. Under appropriate distance, the dissipative coupling between these two remote domain walls is previsible.  This scenario can also be utilized to dissipatively couple different kinds of spin textures, such as chiral magnetic soliton lattice~\cite{Togawa-12,Igor-prb}, skyrmions~\cite{Seki-12,Fert-13}. This coupling could have profound applications in hybrid spintronics structures and devices. Further, we predict that any physical field which can transfer coherence and serve as a dissipative reservoir can be used to sustain dissipative coupling.

\subsection{Expanding the horizon for developing metamaterials}
Coherent coupling has already been utilized for developing metamaterials, leading to for example the demonstration of classical analogues of electromagnetically induced transparency in metamaterials~\cite{Tassin-12}. Such a classical analogue usually applies when the dissipative loss of the radiative resonator is much smaller than the coupling strength. Common wisdom would often consider dissipation processes as a nuisance that may hinder the realization of exotic phenomena in metamaterials. This was highlighted in the earlier debating of whether negative refraction would make a perfect lens~\cite{Pendry-00,Hooft,Pendry-01,Garcia-03}. But in light of the physics of dissipative coupling that origins from the dissipation process, one may consider designing metamaterials that are coupled to waveguides, so that the effect of cooperative damping as we discussed in Section \ref{sec3B} is utilized in developing metamaterials~\cite{Tan-14,Guo-18}. One of the many effects it may lead to is the creation of the ultrasharp resonance, as shown in Fig. \ref{fig:6}, which is induced by the interplay of coherent and dissipative coupling. Such ultrasharp resonances can be used for developing metamaterials based sensors.

\section{Conclusion}
This article describes an emerging novel light-matter interaction in open cavity magnonics systems, which appears as level attraction of the hybridized modes. The word ``open'' here means the external damping rate is much larger than the intrinsic damping rate or even becomes comparable to the cavity eigenfrequency. This coupling mechanism is called dissipative coupling for its dissipation nature. Experimental and theoretical studies of magnon-photon dissipative couplings are reviewed. The canonical dissipative coupling originates from cooperative external dampings or cooperative coupling to a damped mode~\cite{Clerk-15,Yu-19,Du-19,Wang-19,Rao-19-2,Yao-19}, which is genuinely a dissipative phenomenon.

This physics of dissipative coupling has broad relevance. Based on the characteristic feature of the interplay between coherent and dissipative couplings, we have briefly outlined the potential applications for engineering hybrid quantum systems, harnessing dissipative coupling for cavity spintronics, and expanding the horizon for developing metamaterials. The future looks bright for developing both quantum and classical information technologies by harnessing dissipative couplings.

\begin{acknowledgments}
		
This article has been based on collaboration with many colleagues as identified in the reference section. We thank all members and alumni of the Dynamic Spintronics Group at the University of Manitoba, especially L.H. Bai, Y.S. Gui, M. Harder, P. Hyde, J. W. Rao, P.C. Xu, Y. Yang, B.M. Yao, and Y.T. Zhao for their contributions. We are grateful for the communication with G. S. Agarwal, W.E. Bailey, P. Barclay, G.E.W. Bauer, G.S. Beach, N.R. Bernier, I. Boventer, H. Chen, M. Freeman, V.L. Grigoryan, H. Guo, J.A. Haigh, X. Han, B. Heinrich, X.F. Jin, T.J. Kippenberg, L.Q. Liu, W. Lu, R.D. McMichael, A. Metelmann, I. Proskurin, Lu J. Sham, R. L. Stamps, M.D. Stiles, H.X. Tang, O. Tchernyshyov, Y. Tserkovnyak,  M. Weides, K. Xia, J. Xiao, Y. Xiao, A. Yelon, J.Q. You, and T. Yu. This work has been funded by NSERC Discovery Grants and NSERC Discovery Accelerator Supplements (C.-M. H.).
		
\end{acknowledgments}
	
\appendix

\end{document}